\documentclass{nature}
\usepackage{amssymb}
\usepackage{aas_macros}
\usepackage{graphicx}
\usepackage{pdflscape}
\usepackage{hyperref}
\usepackage{txfonts}
\usepackage{textcomp}
\usepackage{threeparttable}
\usepackage{marvosym}
\usepackage{multirow}

\usepackage{amssymb,amsmath}
\usepackage{xcolor}
\usepackage{ulem}
\usepackage{lineno}
\usepackage{pdfpages}
\usepackage[caption = false]{subfig}
\usepackage{caption}
\makeatletter
\usepackage[symbol]{footmisc}
\newcommand{\EXTTAB}[1] {Supplementary Table}
\usepackage{longtable}
\usepackage{ulem}

\def\be{\begin{eqnarray}}
\def\ee{\end{eqnarray}}

\newcommand{\fraction}[3]{\left(\frac{#1}{#2}\right)^{#3}}
\newcommand{\fractionz}[2]{\left(\frac{#1}{#2}\right)}

\let\saved@includegraphics\includegraphics
\AtBeginDocument{\let\includegraphics\saved@includegraphics}
\renewenvironment*{figure}{\@float{figure}}{\end@float}
\makeatother

\title{Decadal evolution of a repeating fast radio burst source}
\begin{document}
\maketitle
\author{
P. Wang$^{1,2}$\href{https://orcid.org/0000-0002-3386-7159}{\includegraphics[scale=0.08]{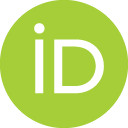}}\footnotemark[1],
J.~S. Zhang$^{1,3}$\href{https://orcid.org/0009-0005-8586-3001}{\includegraphics[scale=0.08]{ORCIDiD.png}}\footnotemark[1],
Y.~P. Yang$^{4}$\href{https://orcid.org/0000-0001-6374-8313}{\includegraphics[scale=0.08]{ORCIDiD.png}}\footnotemark[1],
D.~K. Zhou$^{5}$\href{https://orcid.org/0000-0002-7420-9988}{\includegraphics[scale=0.08]{ORCIDiD.png}},
Y.~K. Zhang$^{1,3}$\href{https://orcid.org/0000-0002-8744-3546}{\includegraphics[scale=0.08]{ORCIDiD.png}},
Y. Feng$^{5,6}$\href{https://orcid.org/0000-0002-0475-7479}{\includegraphics[scale=0.08]{ORCIDiD.png}},
Z.~Y. Zhao$^{7}$\href{https://orcid.org/0000-0002-2171-9861}{\includegraphics[scale=0.08]{ORCIDiD.png}},
J.~H. Fang$^{5}$\href{https://orcid.org/0000-0001-9956-6298}{\includegraphics[scale=0.08]{ORCIDiD.png}},
D. Li$^{8,1,5}$\href{https://orcid.org/0000-0003-3010-7661}{\includegraphics[scale=0.08]{ORCIDiD.png}\textsuperscript{\Letter}},
W.~W. Zhu$^{1,2}$\href{https://orcid.org/0000-0001-5105-4058}{\includegraphics[scale=0.08]{ORCIDiD.png}},
B. Zhang$^{9}$\href{https://orcid.org/0000-0002-9725-2524}{\includegraphics[scale=0.08]{ORCIDiD.png}\textsuperscript{\Letter}},
F.~Y. Wang$^{7,10}$\href{https://orcid.org/0000-0003-4157-7714}{\includegraphics[scale=0.08]{ORCIDiD.png}},
Y.~F. Huang$^{7,10}$\href{https://orcid.org/0000-0001-7199-2906}{\includegraphics[scale=0.08]{ORCIDiD.png}},
R. Luo$^{11}$\href{https://orcid.org/0000-0002-4300-121X}{\includegraphics[scale=0.08]{ORCIDiD.png}},
J.~L. Han$^{1}$\href{https://orcid.org/0000-0002-9274-3092}{\includegraphics[scale=0.08]{ORCIDiD.png}},
K.~J. Lee$^{12,13,1}$\href{https://orcid.org/0000-0002-1435-0883}{\includegraphics[scale=0.08]{ORCIDiD.png}},
C.~W. Tsai$^{1,2,3}$\href{https://orcid.org/0000-0002-9390-9672}{\includegraphics[scale=0.08]{ORCIDiD.png}},
Z.~G. Dai$^{14}$\href{https://orcid.org/0000-0002-7835-8585}{\includegraphics[scale=0.08]{ORCIDiD.png}},
H. Gao$^{2,15}$\href{https://orcid.org/0000-0003-2516-6288}{\includegraphics[scale=0.08]{ORCIDiD.png}},
X.~P. Zheng$^{16}$
J.~H. Cao$^{1,3}$\href{https://orcid.org/0009-0000-7501-2215}{\includegraphics[scale=0.08]{ORCIDiD.png}},
X.~L. Chen$^{1}$\href{https://orcid.org/}{\includegraphics[scale=0.08]{ORCIDiD.png}},
E. G\"{u}gercino\u{g}lu$^{1}$,
J.~C. Jiang$^{1,12}$\href{https://orcid.org/0000-0002-6465-0091}{\includegraphics[scale=0.08]{ORCIDiD.png}},
W.~C. Jing$^{1,3}$\href{https://orcid.org/0000-0002-1056-5895}{\includegraphics[scale=0.08]{ORCIDiD.png}},
Y. Li$^{17}$,
J. Li$^{14}$\href{https://orcid.org/0000-0003-1720-9727}{\includegraphics[scale=0.08]{ORCIDiD.png}},
W.~J. Lu$^{1,3}$\href{https://orcid.org/0000-0001-5653-3787}{\includegraphics[scale=0.08]{ORCIDiD.png}},
J.~W. Luo$^{18}$\href{https://orcid.org/0000-0002-9642-9682}{\includegraphics[scale=0.08]{ORCIDiD.png}},
F. Lyu$^{19}$\href{https://orcid.org/0000-0002-6072-3329}{\includegraphics[scale=0.08]{ORCIDiD.png}},
C.~C. Miao$^{5}$\href{https://orcid.org/0000-0002-9441-2190}{\includegraphics[scale=0.08]{ORCIDiD.png}},
C.~H. Niu$^{16}$\href{https://orcid.org/0000-0001-6651-7799}{\includegraphics[scale=0.08]{ORCIDiD.png}},
J.~R. Niu$^{1,3}$\href{https://orcid.org/0000-0001-8065-4191}{\includegraphics[scale=0.08]{ORCIDiD.png}},
Y. Qu$^{9}$\href{https://orcid.org/0000-0003-4721-4869}{\includegraphics[scale=0.08]{ORCIDiD.png}},
W.~Y. Wang$^{3}$\href{https://orcid.org/0000-0001-9036-8543}{\includegraphics[scale=0.08]{ORCIDiD.png}},
Y.~D. Wang$^{1,3}$\href{https://orcid.org/0000-0002-7372-4160}{\includegraphics[scale=0.08]{ORCIDiD.png}},
Y.~B. Wang$^{20}$\href{https://orcid.org/0000-0002-6592-8411}{\includegraphics[scale=0.08]{ORCIDiD.png}},
C.~J. Wang$^{20}$,
Q. Wu$^{7}$\href{https://orcid.org/0000-0001-6021-5933}{\includegraphics[scale=0.08]{ORCIDiD.png}},
Y.~S. Wu$^{20}$,
S.~M. Weng$^{21}$\href{https://orcid.org/0000-0001-7746-9462}{\includegraphics[scale=0.08]{ORCIDiD.png}},
D. Xiao$^{17}$\href{https://orcid.org/0000-0002-4304-2759}{\includegraphics[scale=0.08]{ORCIDiD.png}},
H. Xu$^{1,12}$\href{https://orcid.org/0000-0002-5031-8098}{\includegraphics[scale=0.08]{ORCIDiD.png}},
J.~M. Yao$^{22}$,
C.~F. Zhang$^{1,12}$\href{https://orcid.org/0000-0002-4327-711X}{\includegraphics[scale=0.08]{ORCIDiD.png}},
R.~S. Zhao$^{23}$\href{https://orcid.org/0000-0002-1243-0476}{\includegraphics[scale=0.08]{ORCIDiD.png}},
Q.~Z. Liu$^{17,24}$,
J. Zhang$^{24}$,
D.~J. Zhou$^{1,3}$\href{https://orcid.org/0000-0002-6423-6106}{\includegraphics[scale=0.08]{ORCIDiD.png}},
L. Zhang$^{1,25}$\href{https://orcid.org/0000-0001-8539-4237}{\includegraphics[scale=0.08]{ORCIDiD.png}},
Y.~H. Zhu$^{1,3}$\href{https://orcid.org/0009-0009-8320-1484}{\includegraphics[scale=0.08]{ORCIDiD.png}}

\makeatletter
\def\thanks#1{\protected@xdef\@thanks{\@thanks
        \protect\footnotetext{#1}}}
\makeatother

\maketitle
\footnotetext[1]{These authors contributed equally to this work.}
\footnotetext{\Letter~ dili@mail.tsinghua.edu.cn; bing.zhang@unlv.edu}
\begin{affiliations}
\item National Astronomical Observatories, Chinese Academy of Sciences, Beijing 100101, China
\item Institute for Frontiers in Astronomy and Astrophysics, Beijing Normal University,  Beijing 102206, China
\item University of Chinese Academy of Sciences, Beijing 100049, China
\item South-Western Institute for Astronomy Research, Yunnan University, Kunming, Yunnan 650504, China
\item Zhejiang Lab, Hangzhou, Zhejiang 311121, China
\item Institute for Astronomy, School of Physics, Zhejiang University, Hangzhou 310027, China
\item School of Astronomy and Space Science, Nanjing University, Nanjing 210023, China
\item Department of Astronomy, Tsinghua University, Beijing 100084, China
\item Department of Physics and Astronomy, University of Nevada, Las Vegas, Las Vegas, NV 89154, USA
\item Key Laboratory of Modern Astronomy and Astrophysics (Nanjing University), Ministry of Education, Nanjing 210093, China
\item Guangzhou University, Guangzhou 510006, China
\item Department of Astronomy, Peking University, Beijing 100871, China
\item Kavli Institute for Astronomy and Astrophysics, Peking University, Beijing 100871, China
\item Department of Astronomy, School of Physical Sciences, University of Science and Technology of China, Hefei 230026, China
\item Department of Astronomy, Beijing Normal University, Beijing 100875, China
\item Institute of Astrophysics, Central China Normal University, Wuhan 430079, China
\item Purple Mountain Observatory, Chinese Academy of Sciences, Nanjing 210023, China
\item College of Physics, Hebei Normal University, Shijiazhuang 050024, China
\item Institute of Astronomy and Astrophysics, Anqing Normal University, Anqing 246133, China
\item Tencent Youtu Lab, Shanghai 200030, China
\item Key Laboratory for Laser Plasmas (MoE), School of Physics and Astronomy, Shanghai Jiao Tong University, Shanghai 200240, China
\item Xinjiang Astronomical Observatory, Chinese Academy of Sciences, Urumqi 830011, China
\item Guizhou Provincial Key Laboratory of Radio Astronomy and Data Processing, Guizhou Normal University, Guiyang 550001, China
\item College of Physics and Electronic Engineering, Qilu Normal University, Jinan 250200, China
\item Centre for Astrophysics and Supercomputing, Swinburne University of Technology, P.O. Box 218, Hawthorn, VIC 3122, Australia
\end{affiliations}
\vspace{0.15in}
\begin{abstract}
The origin of fast radio bursts (FRBs), the brightest cosmic radio explosions, is still unknown \cite{petroff22, kumar23, zhang23}. 
Bearing critical clues to FRBs' origin, the long-term evolution of FRBs has yet to be confirmed, since the field is still young and most FRBs were seen only once. Here we report clear evidence of decadal evolution of FRB~20121102A, the first precisely localized repeater. In conjunction with archival data,  our FAST and GBT monitoring campaign since 2020 reveals a significant $\sim$7$\%$ decline of local dispersion measure (DM). The rotation measure (RM) of 30,755$\pm$16 $\mathrm{rad\,m^{-2}}$ detected in the last epoch represents a 70$\%$ decrease compared to that from December 2016 \cite{Hilmarsson21}. The $\sigma_{RM}$ parameter, which describes the complexity of the magneto-ionic environment surrounding the source, was shown to have decreased by 13\%.
These general trends reveal an evolving FRB environment, which could  originate from an early-phase supernova associated with an enhanced pair wind from the FRB central engine.
\end{abstract}

\linespread{1.15}
Understanding the long-term  evolution of the magnetized environment around repeating FRBs is an important step toward revealing the origin of such cosmic explosions. Following the previous extremely explosive phase in 2019 \cite{li21}, we have carried out a continuous monitoring campaign of the repeating FRB 20121102A since Mar. 2020 with the Five-hundred-meter Aperture Spherical Radio Telescope (FAST \cite{nan11, li18}) at L-band (1-1.5 GHz) and Green Bank Telescope (GBT) at C-band (3.95-5.85 GHz) for a total of 98.3 hours (see Supplementary Table 1). The source was quiescent for 2 years (12 bursts total) in 2020 and 2021 and reactivated in Aug. 2022. We detected a total of 555 independent bursts (see Supplementary Table 2) in two episodes with FAST, i.e. Episode I (from 15 Aug. to 1 Oct. 2022; 431 bursts) and Episode II (from 5 Feb. to 20 Mar. 2023; 124 bursts), respectively (see Methods). The combination of long-term observation, high cadence, and sensitive observations has enabled us to study a significant number of repeating bursts, uncovering previously unseen characteristics of temporal evolution. Fig.~\ref{fig1} shows the observing cadence and burst statistics as a function of time, with the accumulated counts in panel (a) as well as daily average burst counts and rates in panel (b) for Episode I and II. The burst rate peaked at 188~hr$^{-1}$ on 3rd Sep. 2022 (Episode I) and then 135~hr$^{-1}$ on 3rd Mar. 2023 (Episode II), detected over the respective 30 and 20 minutes sessions with FAST, whereas very few bursts can be detected above the thresholds of GBT (L-band, 2$\times$10$^{38}$ erg) or CHIME (8.8$\times$10$^{38}$ erg). During both episodes, the maximum burst rate is higher than the 122~hr$^{-1}$ seen in the 2019 extremely explosive phase \cite{li21}, and the energy distribution of these bursts exhibits no significant temporal evolution  above the threshold of 90$\%$ completeness of the FAST detection.  Most bursts are emitted in a relatively narrow frequency range within the FAST observation band, with a typical emission bandwidth of 243$\pm$48 MHz (see Methods). The emission bandwidth morphology can be quantified as the ratio of burst bandwidth over central frequency, with a typical value of 0.18$\pm$0.03, and shows no temporal evolution during the episodes in panel (d).

The DM and RM provide valuable information about the intervening and circumburst environments. The early-time observation of FRB 20121102A shows $\langle$DM$_{obs}\rangle$=565 pc cm$^{-3}$ and the DM evolution is not significant within the 2$\sigma$ level (see Methods). The structure-optimized DM$_{obs}$ of the bursts are measured using ``\textit{DM-power}" \cite{lin22,nimmo21}, with a mean value of 553$\pm$0.15 ${\rm pc\ cm^{-3}}$ and 552.5$\pm$0.11 ${\rm pc\ cm^{-3}}$ during Episodes I and II, respectively. Based on the redshift of the host galaxy, the intergalactic medium (IGM) contribution of FRB 20121102A is about DM$_{IGM}$=164 pc cm$^{-3}$ (adopting the Planck cosmological parameters and the IGM ionization fraction $f_{IGM}$=0.83 \cite{yangzhang17}), and the Milky Way interstellar medium (ISM) contribution is about DM$_{MW}$ = 218 pc cm$^{-3}$ \cite{tendulkar17}. Subtracting the host galaxy and Milky Way contributions, the local DM contributed by FRB environment is constrained to be DM$_{local}$ = DM$_{obs}$-DM$_{MW}$-DM$_{IGM}$. Fig.~\ref{fig2}b shows the DM temporal evolution of FRB 20121102A over a 10-year period, which reveals a significant decline of DM in Episode I and II by 8–17 $\rm$ pc\ cm$^{-3}$ (or 4–9$\%$ DM$_{local}$) compared to earlier detections \cite{scholz16,petroff16, Oostrum20,cruces21,li21,hewitt22}. Since the DM variations contributed by the IGM and the ISM are extremely small \cite{yangzhang17}, combining all the data, the observed DM evolution should be caused by the plasma local to the FRB source, leading to
\begin{equation}
    \frac{dDM}{dt} \simeq \frac{dDM_{obs}}{dt} =\begin{cases}\ \ 0.86\pm0.78\ pc\ cm^{-3} yr^{-1},\ for\ \rm{MJD}\textless58785,  &\\ -3.93\pm0.11\ pc\ cm^{-3} yr^{-1},for\ 58785\textless \rm{MJD}\textless60064.&\end{cases} 
\end{equation} %
The 1$\sigma$ errors are acquired through an Markov Chain Monte Carlo (MCMC) fit. Assuming the prior tendency is stable with no significant variation, the observed DM decline during the period of 2022-2023 (MJD 58785-60064) holds a statistical significance of 11.8$\sigma$. Combining a nominal velocity along the line of sight (LOS) $v_{LOS}$ with an observed slope of -3.93$\pm$0.11 pc cm$^{-3}$ yr$^{-1}$ implies a spatial scale of the structure responsible for the dDM/dt $\sim$ n$_e v_{LOS}\sim$ -3.9($\pm$0.1)$\times$10$^6$ km/s cm$^{-3}$. Note that if the rapid burst-to-burst fluctuations of DM$_{obs}$ 2-4 $\rm$ pc\ cm$^{-3}$ (Fig.~\ref{fig2}c) are not as a result of intra-burst frequency drift, they should be attributed to the external causes such as turbulent motion of clumps or filaments in the surrounding environment along the line of sight. 

We also perform a search for RM within the range of $-5.0\times10^5$ to $5.0\times10^5$\,$\mathrm{rad\,m^{-2}}$, which encloses the largest value of $\mathrm{RM} \sim 10^5\,\mathrm{rad\,m^{-2}}$ reported in \cite{michilli18} by a wide margin (see Methods). We detect the RM from a bright burst of B489 in the L-band on 3 Mar. 2023, with a linear polarization fraction of 4\% at RM = 30755 $^{+10}_{-15} \mathrm{rad\,m^{-2}}$. The fraction of linear polarization for FRB 20121102A is consistent with depolarization attributable to RM scattering \cite{Feng22}. All previous polarization detections are accomplished at frequency bands higher than the L-band ($\sim$1.4 GHz), for example, at 4-8\,GHz  \cite{michilli18} or at 3-8\,GHz \cite{hi21}. No other significant peak was found on the Faraday spectrum. Fig.~\ref{fig2}a shows the temporal RM variation of FRB 20121102A over 8 years and the RM detection of B489 (see Methods). Combining all the available data, it is clear that the RM of B489 still follows the downward evolution trend as observed between 2016 and 2019 \cite{michilli18, Hilmarsson21, Plavin22}, and drops by $\sim$70\% from the last measurement. The slope of the decay is in the range of -28.5$^{+2.0}_{-1.8}$ rad m$^{-2}$ day$^{-1}$, with a 1$\sigma$ error estimated using an MCMC fit.

The RM scatter of $\sigma_{RM}$ = 26.73$\pm$0.96 rad m$^{-2}$ (see Methods) is shown to have decreased by $\sim$13$\%$ during the 2022-2023 FAST observing campaigns as compared to the previous determinations of $\sigma_{RM}$ = 30.9 rad m$^{-2}$ \cite{Feng22}. This change in the $\sigma_{RM}$ may reflect the evolution of the inhomogeneous magneto-ionic environment. We estimate the lower limit range on the average magnetic field along the LOS, as $\langle B_{\parallel} \rangle \simeq 1.2\left[ \frac{\Delta RM/(10^4\ rad\ m^{-2})}{\Delta DM/(10\ pc\ cm^{-3})} \right]$ mG = $1.2\left( \frac{dRM}{dt} /\frac{dDM}{dt} \right)\bigg|_{MJD>58785}^{MJD<60064}$mG = 2.2 to 2.6 mG, assuming that the DM and RM variations arise from the same fully turbulent medium. As a comparison, the lower limit of $\langle B_{\parallel} \rangle $ is slightly higher than the previous value of 0.6-2.4 mG \cite{michilli18}, but is smaller than that of FRB 20190520B with 3-6 mG \cite{Reshma23}. Spectral depolarization due to high RM and $\sigma_{RM}$ values similar to FRB 20121102A has also been found in some FRBs discovered at low frequencies with CHIME or FAST. It is worth noting that repeating FRB sources with RM detected at high frequencies, such as FRB 20121102A \cite{michilli18} and FRB 20190520B \cite{niu22}, have large RM values and are also associated with persistent radio sources (PRSs). There is a theoretical prediction that the brightness of PRS is proportional to the value of RM \cite{yang22}. Such a prediction has been recently verified by the detection of a faint PRS \cite{bruni24} associated with the repeating FRB 20201124A with a moderate RM \cite{xu22}. 

Peak flux density, pulse width, and fluence of each burst were measured. Given the redshift $z=0.193$ \cite{tendulkar17}, we adopt the latest cosmological parameters from the Planck results \cite{Planck} to derive the corresponding luminosity distance ($D_{\rm L}=949$~Mpc). 
The total isotropic energy emitted by the 555 bursts detected during the episodes is $1.92 \times 10^{40}$ erg. The total energy emitted during the observing campaigns is already $\sim 21\%$ of the total dipolar magnetic field energy of a magnetar ($\simeq 1.7 \times 10^{47}$ erg \cite{li21}) with a surface magnetic field strength $\sim 10^{15}$ G, if a typical radiative efficiency $\eta \sim 10^{-4}$ and a constant burst rate are assumed. The estimation would not change significantly even if the beaming effect is considered \cite{zhang23}. The high energy budget requirement and the high burst rate of this FRB source pose challenges to the magnetar models that invoke a relatively low radio emission efficiency, e.g. the synchrotron maser mechanism \cite{plotnikov19}.

Fig.~\ref{fig3} shows the energy distribution of FRB 20121102A. The energy probability density functions (PDFs) can be well characterized by a Log-Normal (LN) distribution, and the PDFs are the same using the Kolmogorov-Smirnov (K-S) test during the observing campaigns of 2019, 2022, and 2023 (see Methods). The characteristic energies $E_0 \sim 5.8\times 10^{37}$ erg (2019, MJD$\textgreater$58740), $7.9\times 10^{37}$ erg (2022, Episode I), and $6.6\times 10^{37}$ erg (2023, Episode II) are also comparable and robust against uncertainties in detection threshold and choices of pipelines. The emission mechanisms become less efficient below the characteristic energy scale $E_0$. The consistency of the energy PDF and $E_0$ suggests the absence of a temporal evolution  of the central engine. Also, our comprehensive analysis suggests that the burst morphology and year-to-year statistics are both stable in the long term (see Methods). The temporal evolution of DM and RM should be authentic and would therefore be probes of the local environment of the source.

The waiting time between two adjacent bursts is $\delta$t = $t_{\rm i+1}-t_{\rm i}$, where $t_{\rm i+1}$ and $t_{\rm i}$ are the bary-centered arrival times for the ($i+1$)th and ($i$)th bursts, respectively. All waiting times are calculated for bursts within the same observing session to avoid long gaps of days. We normalize the waiting time distribution using peak values, enabling a comparison of the temporal evolution as reflected by the different FAST observing campaigns in 2019/2022/2023.
In Fig ~\ref{fig4}a, the waiting time distributions exhibit a prominent characteristic of a bimodal distribution with two-time scales (milliseconds and tens of seconds) for each of the observing campaigns, each of which can be well fitted by an LN function.
In the 2022-2023 FAST observing campaign, both of the longer timescale peaks are fitted by LN centered at 31$\pm$6 s, indicating an increase in the burst rate relative to the 2019 \cite{li21} peak of 70$\pm$12s. The shorter timescale peaks, which fall within the 3-9 millisecond range, are likely a result of substructure in individual bursts, some of which may be closely spaced independent bursts. This is consistent with the 2019 FAST observing campaign.
The K-S test between the different FAST observing campaigns in 2019/2022/2023, finds a P value of 0.65 (for tens of seconds timescale) and 0.4 (for milliseconds timescale) respectively, indicating that they are statistically the same for each other and have no significant temporal evolutionary properties.
The peaks around tens of seconds timescale in the waiting time distribution are close to the average values for the respective samples (full and high energy). This is also consistent with a stochastic process as in the 2019 FAST observing campaign \cite{li21}.
The distributions of waiting time for 2022 (Episode I, blue) and 2023 (Episode II, red) in three different energy regions are separately shown from top to bottom panels in Fig ~\ref{fig4}b, the longer waiting time peaks for high-energy bursts are attributed to the lower burst rates, and there is no obvious time evolution shown in the waiting time distributions for different energy regions.

No significant periodicity was found (above 3$\sigma$) for either the short timescale (between 1 ms to 600 s) or the longer timescale of 0.028-108 days in the power spectrum (see Methods) during the FAST observing Episodes I and II. The absence of short-timescale periodicity is inconsistent with the expectation of a stably rotating neutron star with a narrow emission beam. Nonetheless, in view that the radio bursts of the Galactic FRB-emitting magnetar SGR 1935+2154 show a wide separation in phase in contrast with bursts \cite{zhu23}, an underlying rotation period cannot be ruled out. The significance of the long-term 157-day activity window cycle \cite{rajwade20} was not greatly enhanced during the full time span, mainly because bursts were not detected during some predicted ``on'' phases. On the other hand, the detected bursts mostly fall into the predicted ``on'' phases (see Methods).  

The decadal evolutions of the DM and the RM of FRB 20121102A shed light on to the environment of active repeating FRB sources. Since the long-term RM evolution is consistent with a monotonic decay, the long-term temporal evolution behaviors of both RM and DM can be mostly dictated by the expansion of the supernova remnant (SNR) ejecta 
\cite{yangzhang17,piro18,hi21,yang23}. 
During a SNR expansion, both the electron density and the parallel magnetic field evolve, leading to a power-law evolution of both the RM and the DM, ${\rm RM}\propto t^{-\alpha}$ and ${\rm DM}\propto t^{-\beta}$.
The decadal observation of FRB 20121102A shows $|d\ln{\rm RM}/dt| = (1/{\rm RM}) |d{\rm RM}/dt| \simeq0.1~{\rm yr^{-1}}$, giving an estimate of the SNR age of $t_{\rm SNR}\sim10\alpha~{\rm yr}$. This result suggests that the associated SNR should be young with an age of a few $\times~(1-10)$ years, consistent with its apparent age of $\gtrsim 12$ years. Such an age corresponds to the free-expansion phase of a young SNR (see Methods). Besides, based on the long-term evolution of both the RM and the DM, we can also estimate the DM contribution from the host interstellar medium to be $\sim200~{\rm pc~cm^{-3}}$ (see Methods).
There is no significant DM change between 2015 and 2019, which deviates from a power-law decay as predicted by the simple model. We propose that the DM decay then could have been offset by the injection of electron-positron pairs from the FRB central engine (see Methods). During the activity phase of a young magnetar, the dissipation of the magnetic energy of the magnetosphere would lead to an enhanced magnetar wind. When the enhanced wind reaches the terminated shock of the magnetar wind nebula, the injection of pairs would cause an extra DM contribution but cause no contribution to the RM. We find that in order to contribute the extra DM of $\Delta{\rm DM}\sim{\rm a~few~pc~cm^{-3}}$ compared with the measurements in 2012 \cite{Spitler2014} and after 2020, the luminosity of the enhanced wind is required to be amplified by hundreds of times compared to that of the pre-active persistent wind. 
These general trends reveal an unraveling FRB environment, which can be well
explained by an early-phase supernova explosion. We expect more evidence for the long term DM and RM evolution of other active repeaters to become available in the coming years, potentially further strengthen the young magnetar scenario as FRB origin.

\bibliographystyle{naturemag}

\begin{addendum}
 \item This work is supported by National Natural Science Foundation of China (NSFC) Programs (No. 11988101, 11725313, 11690024, 12041303, 12473047, U1731238, 12233002, 12375236, 12135009, 12203045); by CAS International Partnership Program (No. 114-A11KYSB20160008); by CAS Strategic Priority Research Program (No. XDB23000000); and the National Key R\&D Program of China (No. 2017YFA0402600, 2021YFA0718500); and the National SKA Program of China (No. 2020SKA0120200, 2022SKA0130100, 2020SKA0120300).
 D.L. is a New Cornerstone investigator.
 P.W. acknowledges support from the CAS Youth Interdisciplinary Team, the Youth Innovation Promotion Association CAS (id. 2021055), and the Cultivation Project for FAST Scientific Payoff and Research Achievement of CAMS-CAS.
 Y.P.Y. acknowledges the support from the ``Science \& Technology Champion Project'' (202005AB160002) and from two ``Team Projects'' – the ``Innovation Team'' (202105AE160021) and the ``Top Team'' (202305AT350002), all funded by the ``Yunnan Revitalization Talent Support Program''.
 Y.F. acknowledges the support from the Leading Innovation and Entrepreneurship Team of Zhejiang Province of China (No. 2023R01008), and Key R\&D Program of Zhejiang (No. 2024SSYS0012).
 This work made use of data from FAST, a Chinese national mega-science facility built and operated by the National Astronomical Observatories, Chinese Academy of Sciences. 
\item[Author Contributions] 
P.W. and J.S.Z. led the FAST data analysis, Y.F. and Y.K.Z. participated in polarization data analysis.
D.L., B.Z., and W.W.Z. are the conveners of the project, coordinated the science team, and launched the FAST observational campaign on FRB~20121102A. 
D.K.Z., K.J.L., C.H.N., C.C.M., J.C.J., H.X., C.F.Z., J.H.C., W.J.L., Y.D.W., J.W.C., J.R.N., Q.W., Y.M.Y. R.S.Z. J.Z., L.Z., D.J.Z., Y.B.W. participated in the FAST data analysis. 
Y.P.Y. leds the theoretical interpretation of DM/RM evolution, Z.Y.Z., F.Y.W., S.M.W., D.K.Z, J.H.F, Y.Z.Z, Y.F.H., L.J., B.Z., Z.G.D., D.X., X.P.Z, E.G., Y.L., J.W.L., X.L.C. and Y.H.Q. contributed on simulation and writing of theoretical implications of observation results.
P.W., D.L., B.Z., Y.P.Y., Z.Y.Z., D.K.Z. and J.H.F. contributed to the writing of the manuscript.
All authors discussed the contents and form the final version of the paper.
\item[Competing Interests] The authors declare that they have no competing financial interests.
\end{addendum}
\begin{figure*}
    \centering
    \includegraphics[scale=0.65]{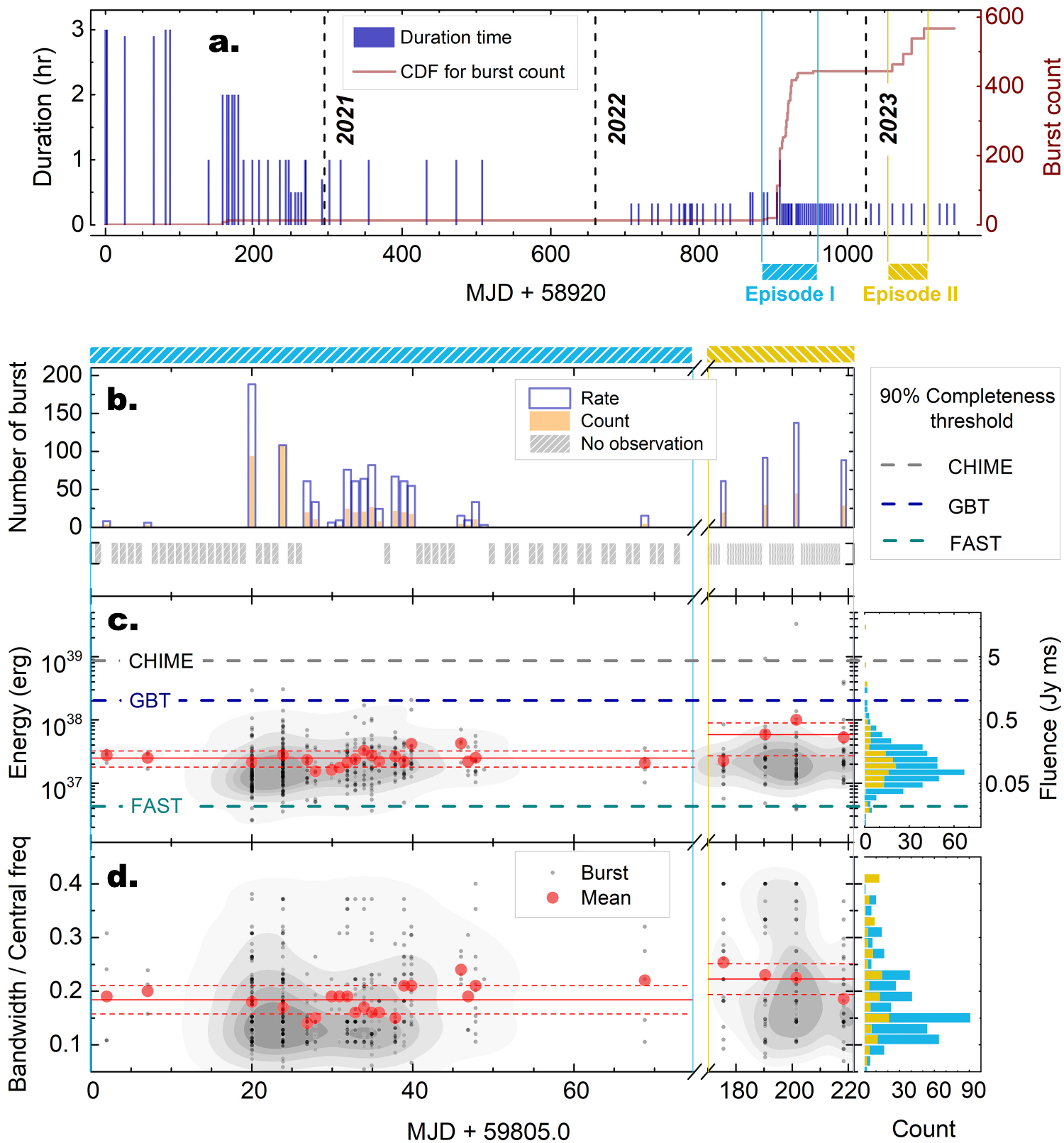}
    \caption{{\bf Timeline, the detected bursts, the temporal energy, and the burst bandwidth distribution during the observing campaign.} Panel a: the duration of each observing session (blue bar) and the cumulative number distribution of the bursts (red solid line). Panel b: the rate (blue bar) and count (orange bar) of the bursts detected during the observing Episodes I and II. The gray bars are days without observations. Panel c and d: time-dependent burst energy and bandwidth distribution, along with their histograms. The black dots are all the 555 bursts, the red dots represent the average value for each observing session. The grey shade represents the 2D Kernel Density Estimation (KDE) of the bursts.}
    \label{fig1}
\end{figure*}
\begin{figure*}
    \centering
    \includegraphics[scale=0.6]{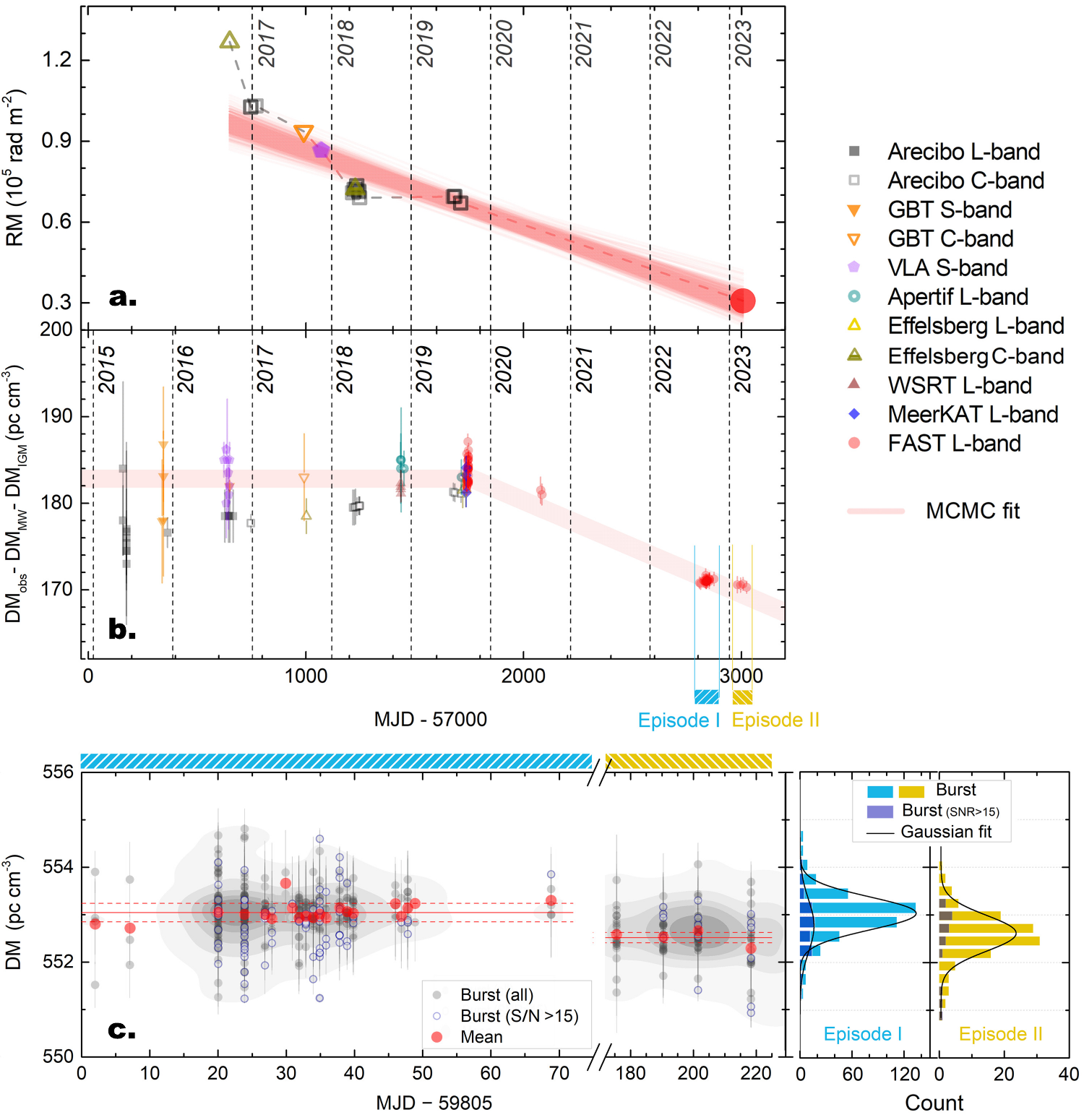}
    \caption{{\bf The RM \& DM temporal evolution of FRB 20121102A.} Panel a: the temporal RM variation for FRB 20121102A over an 8-year period. The red lines show the result of the MCMC fit assuming linear RM evolution over time with a slope of -28.48$^{+1.98}_{-1.76}$ rad m$^{-2}$ day$^{-1}$, while the shaded region indicates 1$\sigma$ statistical error. Panel b: the decadal variation of the DM. The red lines denote the linear fit with a slope 0.86$\pm$0.78 and -3.93$\pm$0.11 pc cm$^{-3}$ yr$^{-1}$ for MJD$\textless$58785 and MJD 58785-60064, respectively. Panel c provides a more detailed view and histogram of the panel (b) during the FAST observing campaigns in 2022-2023 (Episode I and II). Circles and grey dots indicate all 555 bursts and those with a signal-to-noise ratio (S/N) greater than 15, the red dots represent the average value for each observing session. The shaded regions are the KDEs of the bursts. The solid and dashed lines represent the mean and the standard deviation of DM for each Episode, respectively.}
    \label{fig2}
\end{figure*}
\begin{figure}
     \centering
    \includegraphics[scale=0.4]{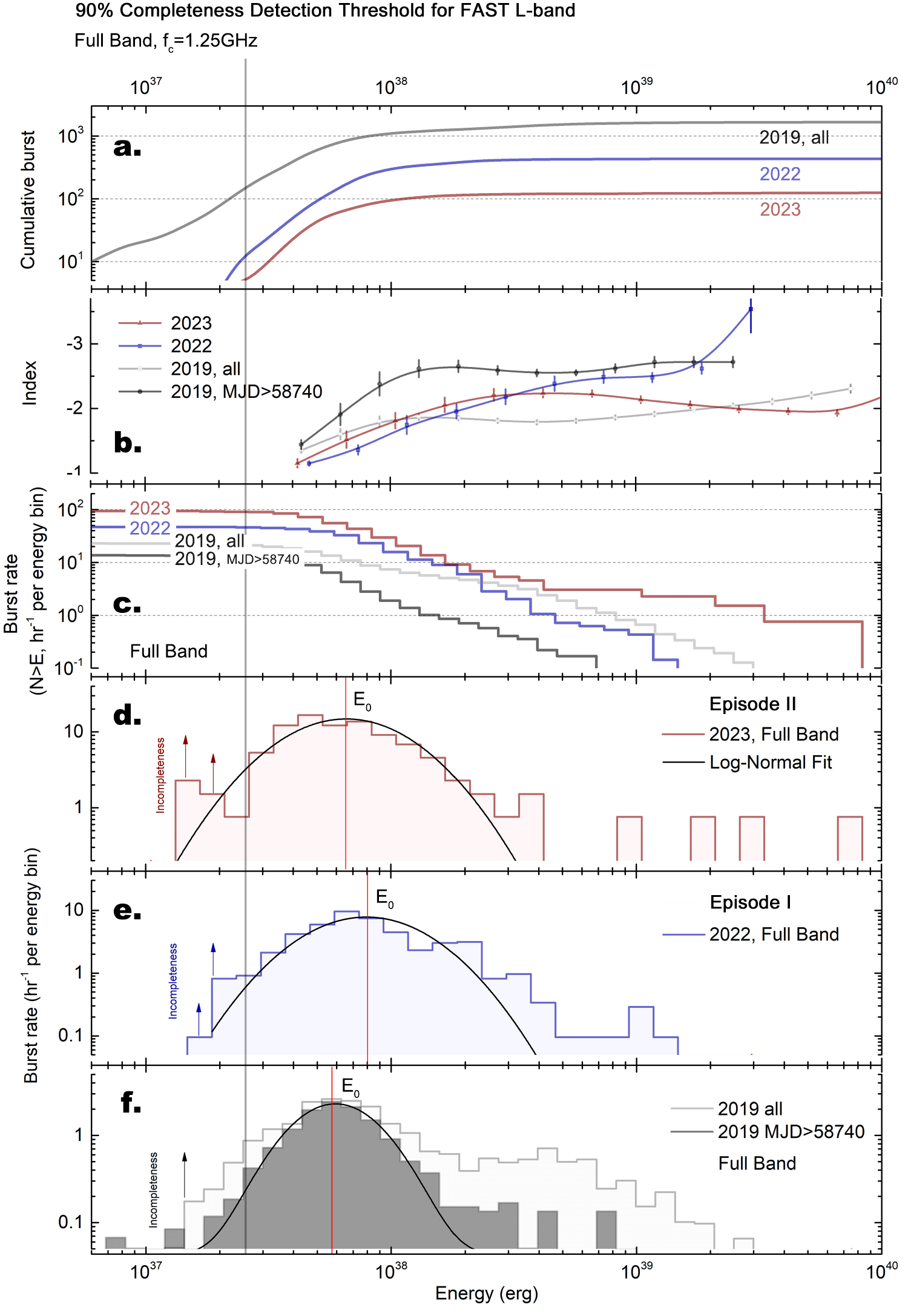}
    \caption{{\bf The energy distribution of FRB 20121102A.} Panel a: the cumulative distributions of the detected bursts during the FAST observing campaigns in 2019, 2022, and 2023. Panel b: the spectral index of single-power law fitting as a function of energy threshold. Panel c: the cumulative energy distributions of FRB 20121102A. Panel d-f: the specific energy distributions for FAST observing campaigns in 2023, 2022, and 2019, separately, from top to bottom. The red, blue, and grey bars (including two cases: all data and MJD $\textgreater$ 58740 only) indicate the energy distribution in the full bandwidth definition, i.e. $\Delta \nu$ = 500 MHz in panels d, e, and f, respectively. The solid black lines result from the Log-Normal (LN) fit, and the red vertical lines indicate the locations of the maximum value in the LN fit. The 90$\%$ detection completeness threshold is shown by the vertical grey solid and dotted lines, corresponding to E$_{90}^{Full-band}$ = 2.5 × 10$^{37}$ erg for an assumed pulse width of 3 ms and scaled as the square root of the pulse width.}
    \label{fig3}
\end{figure}
\begin{figure}
     \centering
    \includegraphics[scale=0.6]{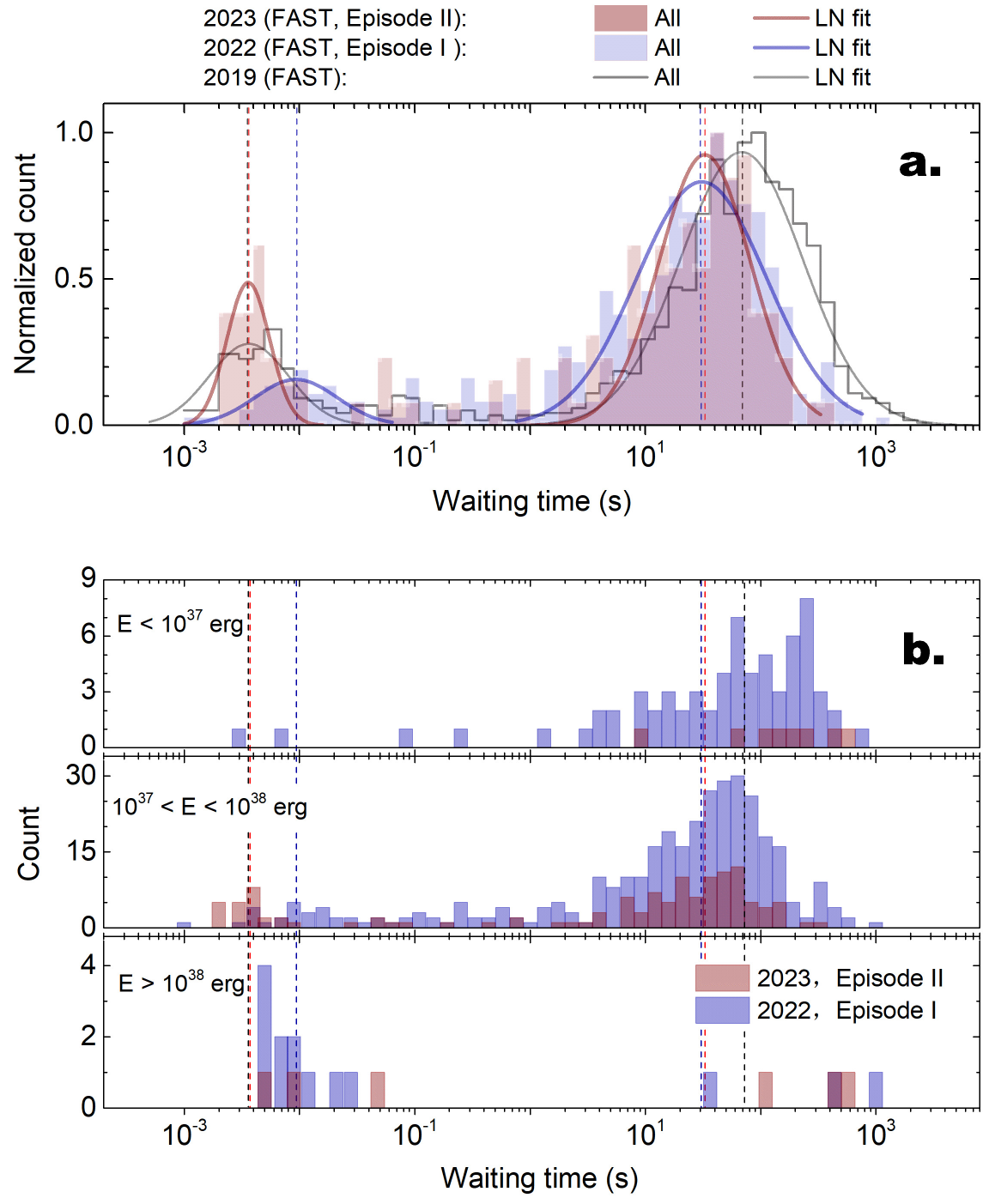}
    \caption{{\bf The waiting time distributions of the bursts.} Panel a: the gray, blue, and red bars denote the normalized waiting time histograms of FAST observations in 2019, 2022, and 2023, respectively, and their corresponding solid lines are the results of LN fit. The dashed lines indicate the locations of the maximum value in LN fit. In the 2022-2023 FAST observing campaign, both of the longer timescale peaks around 31 s, demonstrating an increase in the burst rate compared to the 2019 \cite{li21} peak of 70 s, while the shorter timescale peaks in the range of 3-9 milliseconds, which is also consistent with 2019. Panel b: the distributions of waiting time for 2022 (Episode I, blue) and 2023 (Episode II, red) in three different energy regions are shown from top to bottom.}
    \label{fig4}
\end{figure}
\clearpage
\newpage

\begin{methods}
\setcounter{figure}{0}
\captionsetup[figure]{labelfont={bf},labelformat={default},labelsep=period,name={Extended Data Fig.}}

\section{Observation Campaign and Data Reduction}
FAST has monitored FRB 20121102A since 2019. 
Following an extremely active phase in 2019 \cite{li21}, we perform a combination of FAST (L-band) and GBT (C-band) long-term monitoring of the repeating FRB 20121102A, with a total of 98.3 hours of observations (see Supplementary Table 1 for the full catalog). 
The GBT carried out 10 observation sessions between 12 May and 25 Oct. 2022, using a C-band receiver (3.95-5.85 GHz) for a total of 22 hours.
No bursts were detected during the GBT observation period.
FAST conducted a total of 76.3 hours of observation mainly in the following three sessions: (i) from 12 Mar. 2020 to 29 Jul. 2022 (UTC); (ii) from 1 Aug. 2022 to 18 Dec. 2022 (UTC); (iii) from 7 Jan. 2023 to 30 Apr. 2023 (UTC), through the FRB key project of FAST.
No pulsed emission was detected in the session (i) spanning two years, while we detected a total of 555 bursts in the sessions (ii) and (iii). In this work, we concentrate on an analysis of the active windows: Episode I (MJD 59806-59853) and Episode II (MJD 59980-60023) where the bursts were detected in the sessions (ii) and (iii), respectively.

The FAST L-band receiver has a center frequency of 1.25 GHz, spanning from 1.0 to 1.5 GHz, including a 50 MHz band edge on each side. 
The average system temperature is approximately 25 Kelvin. The recorded FAST data stream for pulsar observations is a time series of total power per frequency channel, stored in the PSRFITS format \cite{hotan2004} from a ROACH-2 based backend, which produces 8-bit sampled data over 4-k frequency channels at 49.152 $\mu$s cadence.

We carefully search for radio pulsations with either a dispersion signature or instrumental saturation in all the data collected during the FAST and GBT observing campaigns. Three types of data processing were performed: (i) dedicated single-pulse search, (ii) baseline saturation search, and (iii) Artificial Intelligence (AI) aided blind search.

(i) Dedicated single-pulse search:
We create the de-dispersed time series for each pseudo-pointing over a range of DMs of 520-600 pc cm$^{-3}$, which should cover all the uncertainties in a semi-blind search. The step size between subsequent trial DMs ($\Delta$DM) is chosen such that over the entire band t($\Delta$DM) = t$_{channel}$. This ensures that the maximum extra smearing caused by any trial DM deviating from the source DM by $\Delta$DM is less than the intra-channel smearing. We use the above dedicated search scheme to de-disperse the data, using the PRESTO toolkits \cite{Ransom2001}.
Then, we use 14 grids (logarithmic space from 0.1 ms to 30 ms) of matching filters to refine the detected S/N. A zero-DM matched filter is also applied to mitigate radio frequency interference (RFI) in blind-search. All the potential candidate plots are subsequently inspected visually. Most of the candidates are identified as RFIs, and the 555 bursts with dispersive signatures are eventually detected with an S/N $\textgreater$6 from the FAST data stream. A fraction of the bursts exhibit a complicated time-frequency morphology.

(ii) Saturation search:
We realize that the receiver of FAST or GBT would be saturated when the radio flux density is as high as hundred-Jansky to mega-Jansky. We therefore search for the possible saturated signals in the recorded data streams. We try to search for the epoch in which 50$\%$ of the frequency channels satisfy the following states: a) the channel is fully saturated (255 value in 8-bit channels), b) the channel is zero-valued, c) the Root Mean Square (RMS) of the bandpass is less than 2. We have not detected any signs of saturation that last longer than 0.3 seconds, thereby ruling out any saturation linked to the observation campaigns.

(iii) Artificial Intelligence (AI) aided blind search:
We have developed a learning-based filtering framework to facilitate efficient blind search. In this framework, we utilize ResNet-18 \cite{he2016deep} as the classification model to convert the input data array into a normalized confidence vector that shows whether the input array belongs to potential candidates. We take the FRB 20121102A data set from the extremely active phase in 2019 \cite{li21} to train the classification network using the standard cross entropy as the training loss factor, and adopt SDM's active learning strategy \cite{xie22} for data processing.

\section{DM Variation}
The DM measurements for each burst were obtained by maximising the structure in the frequency-integrated burst profile, calculated using the \textit{`DM-Power'} algorithm (https://github.com/hsiuhsil/DM-power). For bursts that have obscured morphological structure in the time-frequency dynamics spectrum or low S/N ratios, the \textit{`DM-Power'} algorithm may cause overfitting, and for these bursts, the daily mean DM values are used instead.

To study the reliability of this DM variation tendency, we divided the DM measurement into three time periods based on the dates of the event and the telescopes, i.e. (period I: MJD=56233, period II: 57159$\leqslant$MJD$\leqslant$58785, and period III: MJD$\geqslant$58785), and generated simulated DM values within each period based on the mean and standard deviation of the measured DMs in the respective periods. Period I is the date of discovery \cite{Spitler2014}, its DM measurement was significantly lower than that of the subsequent detections, and the statistical and systematic errors may be underestimated due to the absence of a multi-pulse component.

In a realistic case of ignoring the period I measurement, the null hypothesis was then tested based on the MCMC-generated DMs under the assumption that the DM does not change through time. A slope value was fitted to the each set of generated DMs. The histogram distribution of the DM slopes (57159$\leqslant$MJD$\leqslant$58785, peirod II) were shown in Extended Data Figure ~\ref{EDfig1}, the optimal value of $dDM/dt$ was 0.87 pc cm$^{-3}$ yr$^{-1}$, the uncertainty of measurement was 1$\sigma$ = 0.46 pc cm$^{-3}$ yr$^{-1}$ based on 30,000 trials. Thus, the fitted DM growth rate of $dDM/dt$ = 0 pc cm$^{-3}$ yr$^{-1}$ (red solid line) could result from a null hypothesis sample within 2$\sigma$ significance. From a technical perspective, the lack of the DM-measured temporal evolution in period II is more realistic, whereas the observed DM decline during period III (MJD 58785-60064) has a statistical significance of 11.8$\sigma$, which reveals a confidence evolution characteristic of the DM during 2022-2023.

\begin{figure}
     \centering
    \includegraphics[scale=0.4]{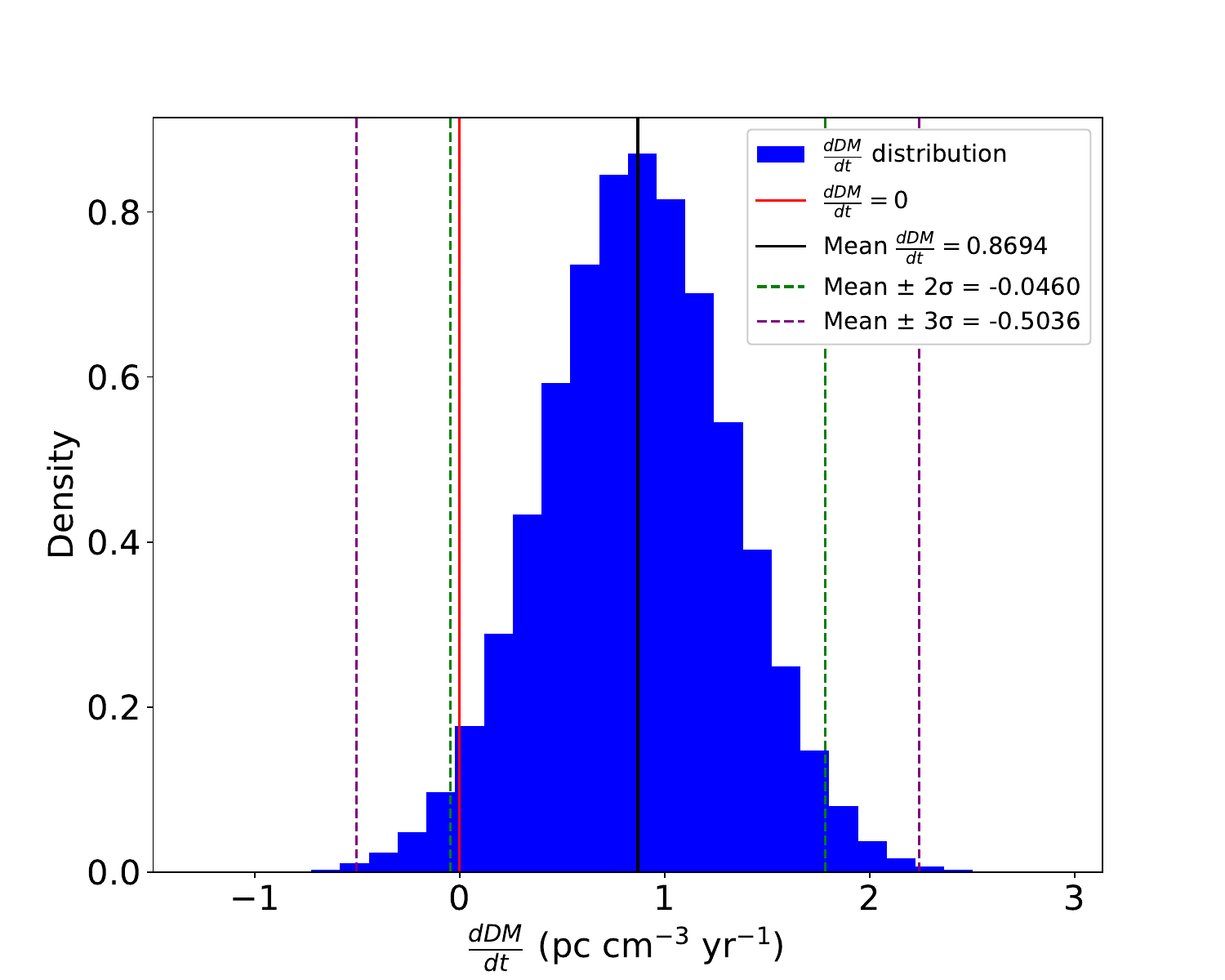}
    \caption{{\bf DM slope distribution of null hypothesis test for 57159$\leqslant$MJD$\leqslant$58785.} The black, green, and purple vertical lines indicate the locations of the mean, 2$\sigma$, and 3-$\sigma$ values in the slope distribution, respectively. The $dDM/dt$ = 0 pc cm$^{-3}$ yr$^{-1}$ is shown by the red solid vertical line, within the 2-$\sigma$ significance compared with the previous detections in period II.}
    \label{EDfig1}
\end{figure}

\section{Burst Rate and Energy Distribution}
To attain a high-quality flux density and polarization calibration solution, a 1K equivalent noise calibration signal was injected before each session and used to scale the data to $T_{\rm sys}$ units. The RMS of off-pulse brightness is constant within $6\%$ for all observations. 
For the FAST L-band observations, the variability in the off-pulse level arises predominantly from the telescopic gain's dependence on the zenith angle. The observatory supplied a zenith angle-dependent gain curve through quasar measurements \cite{jiang19}, which was used to convert Kelvin units to mJy, and applied for each burst.

Without knowledge of the geometric beaming information of FRBs, we calculate the isotropic equivalent burst energy with the following equation of Ref. \cite{zhang18a, gourdji19}
\begin{equation}
E = (10^{39} {\rm erg})\frac{4\pi}{1+z} \left(\frac{D_L}{10^{28}{\rm cm}}\right)^{2}
\left(\frac{F_{\nu}}{\rm Jy\cdot ms}\right)
\left(\frac{\Delta\nu}{\rm GHz}\right),
\end{equation}
where $F_{\nu}$ = $S_{\nu}\times W_{\rm eq}$ is the specific fluence in units of ${\rm erg \ cm^{-2} Hz^{-1}}$ or $\rm Jy\cdot ms$, 
$S_{\nu}$ is the peak flux density which has been calibrated with the noise level of the baseline, and then measured the amount of pulsed flux above the baseline, and $\Delta\nu$ is the bandwidth of each burst. The optimized estimates depend on the spectral shape of the burst. If the FRB spectra are narrow-band with emission within the observing band, it is appropriate to use the observing bandwidth (BW) as $\Delta\nu$ to obtain the fluence/energy. On the other hand, if the FRB spectra are broad-band with emission extending beyond the observing band, $\Delta\nu$ would be more appropriate to use the full-band of the receivers. $W_{\rm eq}$ is the equivalent burst duration, and the luminosity distance $D_L$ = 949 Mpc  corresponds to a redshift $z$ = 0.193 for FRB 20121102A \cite{tendulkar17}.

In the Supplementary Table 2, we adopt two choices of BW (observed BW and full BW of 500 MHz) to calculate $S_{\nu}$ and compare the derived isotropic equivalent burst energies during the FAST observing campaigns in 2019, 2022, and 2023. Although the specific energies obtained using the observed BW are generally lower than those obtained assuming the full BW, there is a similarity in the resulting energy distributions between the two scenarios. After aligning the two energy distributions using cross-correlation, the P-value from the Kolmogorov-Smirnov (K-S) of the energy distribution calculated with two types of BW is 0.83 and 0.98 for the FAST observing campaigns in 2022 (Episode I, panel e) and 2023 (Episode II, panel d), respectively. The normal method of calculating the energy Probability Density Function (PDF) using the full BW is reliable as the BW has no significant effect on the energy distribution above the 90$\%$ detection completeness threshold of E$_{90}^{Full-band}$ = 2.5 × 10$^{37}$ erg and E$_{90}^{Obs-band}$ = 4.5 × 10$^{36}$ erg, respectively, for an assumed pulse width of 3 ms and scaling as the square root of the pulse width \cite{li21}.
Therefore, we uniformly use the full BW fluence for the energy calculation in the subsequent analysis.

The energy/luminosity function can shed light on the emission mechanism of the repeating FRB. We then compare the temporal evolution of the energy distribution of FRB 20121102A shown in panels d, e, and f, which were observed during the 2019, 2022, and 2023 campaigns. The energy PDFs show that weaker bursts in each panel are characterized by the Log-Normal (LN) function, which may become less efficient below the numerically similar characteristic energy scales of $E_0 \sim 5.8\times 10^{37}$ erg (2019, MJD$\textgreater$58740), $7.9\times 10^{37}$ erg (2022, Episode I), and $6.6\times 10^{37}$ erg (2023, Episode II). 
The K-S test of the observing campaigns indicates that the P value for all the pairings is approximately 0.2, which cannot reject the hypothesis that the energy distributions are consistent with each other. This indicates that although sporadic high energy ($\textgreater$3×10$^{38}$ erg) bursts have been observed during the 2022-2023 sessions, the energy distribution does not exhibit a statistical characterization resembling the bimodality of the 2019 time-dependent evolution, implying that the emission modality during the 2019 extremely active phase is unique compared to the 2022-2023 observations. 
 
For a power law distribution of energy $dN/dE = A E^\alpha$, the cumulative energy distribution can be calculated as
\begin{equation}\label{eq:de}
    \begin{aligned}
    N(>E)&=\int_E^{E_{\rm max}} dN = \int_E^{E_{\rm max}} AE^{\alpha} dE\\
    &= \frac{A}{-\alpha-1}\left[E^{\alpha+1}-E_{\rm max}^{\alpha+1}\right]=\frac{A}{-\alpha '}\left[E^{\alpha '}-E_{\rm max}^{\alpha '}\right],
    \end{aligned}
\end{equation}
where $E_{\rm max}$ is the maximum energy for a certain sample, and $\alpha '=\alpha+1$. One can see that in general, the cumulative energy distribution is also roughly a power law with index $\alpha ' < 0$. The index $\alpha '$ has different values for different types of astrophysical sources. We exclude the bursts that are below the 90\% detection threshold (E$_{90}^{Full-band}$ = 2.5×10$^{37}$ erg) and fit the spectral index $\alpha '$ as a function of energy threshold using a single-power law in Extended Data Figure ~\ref{fig3}b. The uncertainties of the fitting are obtained by the Monte Carlo (MC) method by taking the averaged value of the spectral index fitting, 1000 random samples according to the variance of the Gaussian distribution for the cumulative energy distribution. By applying a differing energy threshold and repeating the fitting process, the index $\alpha '$ of a single power law fitting fails to converge to a singular value in panel B. This result suggests that a single power law distribution is not sufficient to describe the cumulative burst energy distribution of FRB 20121102A.

\section{Time-Domain Analysis}
\subsection{Active Periodicity and Time Difference Algorithms}
A plausible active period of about 157 days has been suggested for FRB 20121102A \cite{rajwade20}. We correct the arrival time (ToA) of each burst to an infinite frequency and then convert the ToAs to the solar system barycentre using the DE405 ephemeris. The activation of FRB 20121102A is generally consistent with the proposed periodicity \cite{rajwade20, cruces21} with an S/N = 5.7. However, FAST detected no bursts in a few predicted active time windows.
Combining the bursts collected in Ref.\cite{rajwade20, cruces21, li21} and those newly detected by the long-term monitoring during the 2022-2023 FAST observing campaigns, we obtain a new best-fit period of 156.9 days and a duty cycle of about 62$\%$ (i.e. a 97.3-day on-phase window in the putative period, Extended Data Figure ~\ref{EDfig2}). Alternatively, considering all the non-detections after including the projected turning-on/off time in reality, the S/N of the periodicity drops to 4.3, indicating that the putative period of the source becomes less robust. Such a long-term periodicity may be accommodated within the framework of binary \cite{ioka20} or precession \cite{zanazzi20} models.
\begin{figure*}
    \centering
    \includegraphics[scale=0.85]{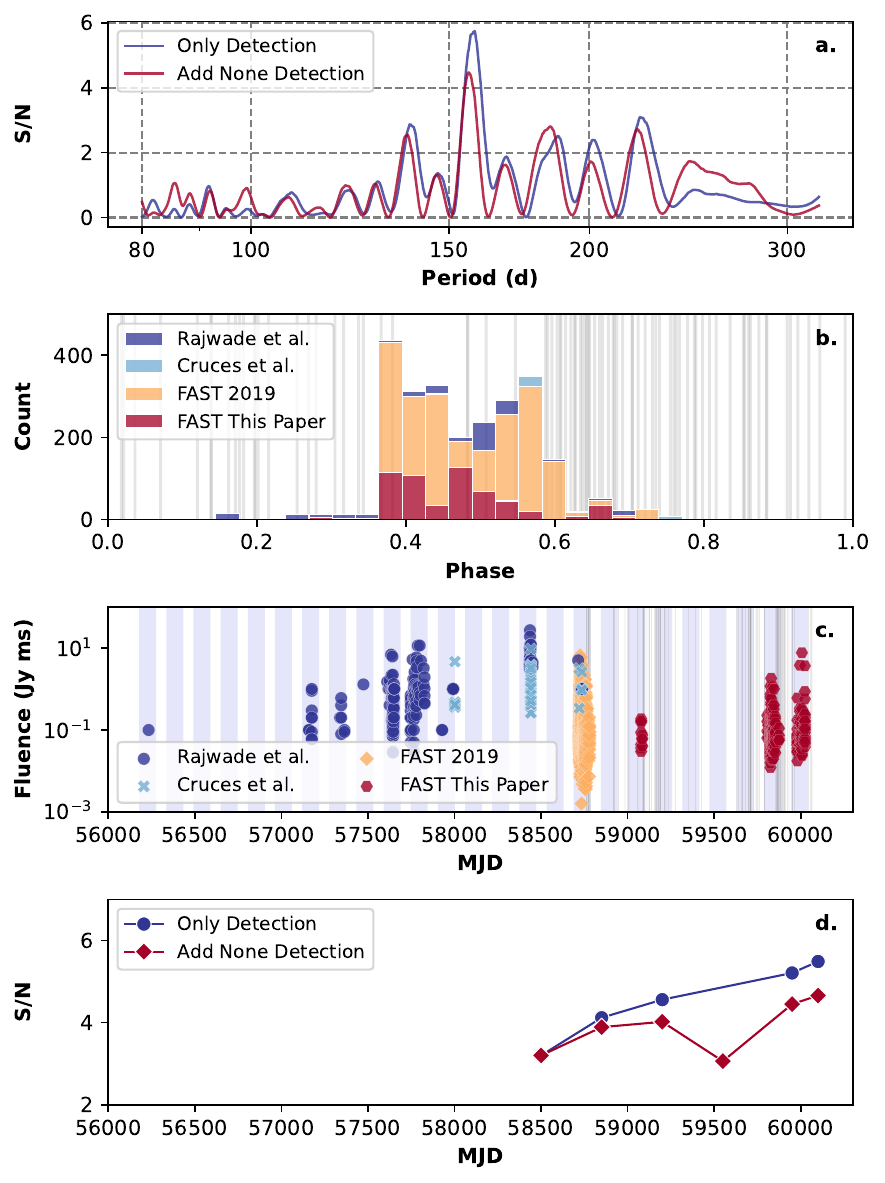}
    \caption{{\bf Periodicity of FRB 20121102A}. Panel a: S/N of the period signal obtained using the Lomb-Scargle Periodogram (LSP). The blue line represents applying the LSP to detected bursts, while the red line represents applying the LSP to the detected bursts plus the observations without bursts detected by FAST. Panel b: burst MJDs folded with a period of 156.9 days, with the gray vertical lines indicating observations with no detected bursts. Panel c: burst MJDs as a function of the fluence. The purple region represents the active window within one period. Panel d: S/N of the period signal at different time nodes.}
    \label{EDfig2}
\end{figure*}
Time difference algorithms (TDAs) are widely used to search for periodic signals from high-energy gamma-ray pulsars \cite{2006ApJ...652L..49A,2008ApJ...680..620Z,2008Sci...322.1218A}. 
This kind of signal is characterized by very sparse events (a few events per day), and the detection of repeating FRBs has similarities. Denoting the sequence of time of arrival (ToA) of each burst as $\{t_i,\space i=1,2,3... \}$, assuming that the sequence is periodic, but the period varies linearly with time, one can use the following equation to correct for the effect of the first derivative of the frequency:
\begin{equation}
    t_i = \tilde{t_i}+\frac{1}{2}\frac{f_1}{f_0}\tilde{t_i}^2,
\end{equation}
where $f_0$, $f_1$ are the frequency and the first derivative of the frequency, and $t_i$, $\tilde{t_i}$ are the ToAs after and before the correction, respectively. The TDA can then be used to search for the period of this sequence. This is done by calculating the time difference between the arrival times, i.e., calculating the difference between each ToA and its following ToA within the window $T_w=T_{obs}/N_w$ for a suitably chosen $N_w$ (this window determines the lower limit of the frequency that can be searched) and then obtaining the time difference sequence, after which the amplitude spectrum of the time difference sequence is calculated using the Fast Fourier Transform (FFT). This approach is effectively equivalent to the following equation:
\begin{equation}
    D_l^{N_w}=\sum_{m=k+1}^{M}\sum_{k=0}^{N-1}a_ka_m e^{-i2\pi l(m-k)/N}, \space l=0,1,...,N/2,
\end{equation}
where $a$ denotes the histogram of the ToA sequence, $N$ denotes the number of points in the histogram, and $M={\rm min}(N-1,k+N/N_w)$ is the upper limit of the outer summation. The periodic signal is found by searching for the peaks of ${\rm Re}(D_l^{N_w})$. When the time window $T_w$ is equal to $T_{obs}$, ${\rm Re}(D_l^{T_w})$ is equivalent to the power spectrum calculated directly using the FFT, but if $T_w$ is less than the observation duration, it cannot be interpreted as a power spectrum. However, we still use the term power to denote it. We use this algorithm to search for $f_0$ and $f_1$ of the ToA sequence. For computational reasons, the search range for $f_1$ was set to $-10^{-6}\sim 10^{-9}$, yielding a total of about 100,000 (related to the specific data set, see below) $\frac{f_1}{f_0}$ ratios to correct for the effect of the first derivative of the frequency.

\subsection{Short Timescale Period Search}
Considering the possibility that the period of the ToA sequence depends on the burst energy, we group a total of 555 bursts according to their energies. The bursts are divided into three groups, and each group contains a sufficient number of ToAs. The first group, named E I, contained 207 ToAs with energy less than $\rm{E} \textless 1. 5\times 10^{37}$ erg; the second group, E II, included 206 ToAs with energy between $1.5\times 10^{37}$ erg and $3\times 10^{ 37}$ erg. Finally, the third group, E III, consisted of 142 ToAs with energy greater than or equal to $\rm{E}\geq 3\times 10^{37}$ erg. This grouping allows each dataset, including the total energy interval ($E_{\rm tot}$), to be analyzed independently.

The stacked FFT (SFFT) algorithm is the classical algorithm for detecting periodic signals \cite{Welch1967TheUO}. This algorithm divides the signal into many equal segments, then performs the FFT algorithm on the data of each segment to obtain the power spectrum, and finally sums these power spectra to obtain the high signal-to-noise power spectrum. We use the SFFT algorithm to search for the short timescale periods of the ToA series and to ensure the robustness of the results, the results are confirmed by a double check of the TDA processing.

For the short timescale periodicity search, considering that the minimum duration of the continuous observations is approximately 1200 seconds, to ensure that even in the case of the longest period, data with the shortest observation time still contains nearly two periods of signal, we set the window size to 600 seconds (resulting in a minimum search frequency of about 0.0017 Hz). Additionally, taking into account that there are only two events with a time difference between consecutive ToAs of less than 7 milliseconds, we set the upper limit for the search frequency to 70 Hz.

\begin{figure*}
	\centering
	\subfloat{\includegraphics[scale=0.4]{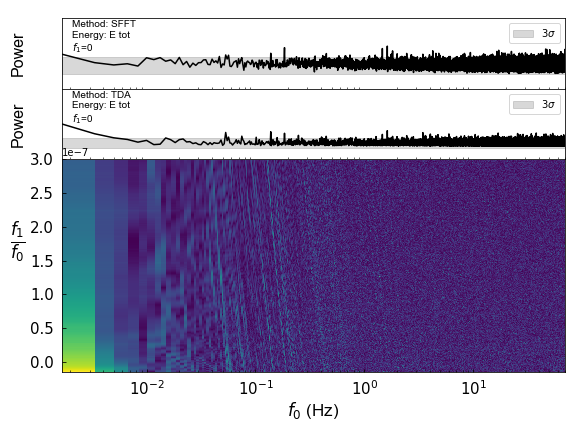}}
	\subfloat{\includegraphics[scale=0.4]{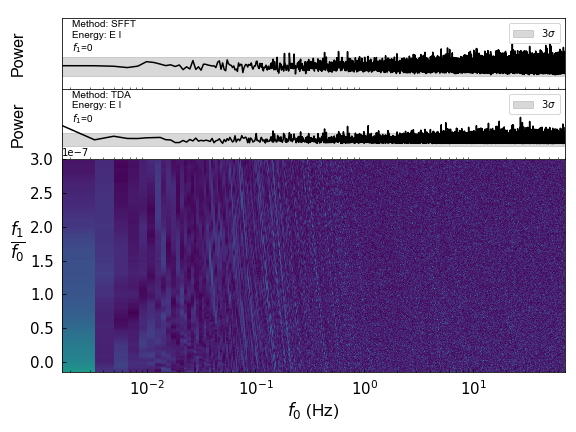}}\\
         \subfloat{\includegraphics[scale=0.4]{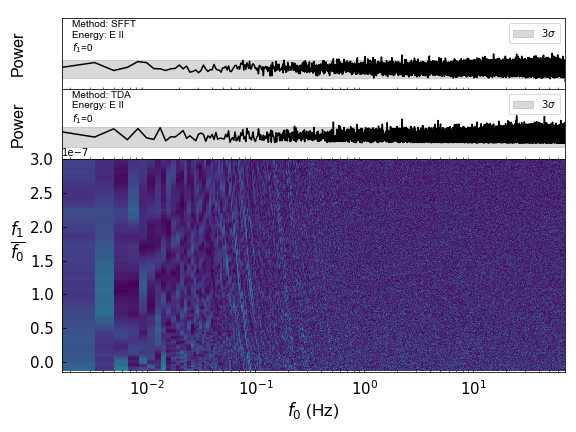}}
         \subfloat{\includegraphics[scale=0.4]{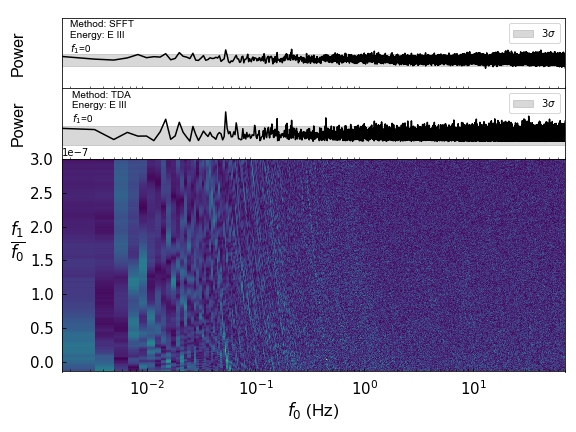}}
   	 \caption{{\bf The results of the search for short timescale periodicity using the TDA and SFFT algorithms.} Each subplot represents a data set of energy intervals. The upper and middle panels in each sub-panel represent the power spectra obtained by the two algorithms for the frequency first-order derivatives at zero, respectively. The heatmaps in the bottom sub-panels represent the result of the TDA search on the first-order derivatives of frequency. The grey regions in the panels mark the 3$\sigma$ region (assuming a Gaussian distribution).}
	 \label{EDfig3}
\end{figure*}
The results of the periodicity search using the TDA and the SFFT algorithms are shown in Extended Data Figure \ref{EDfig3}, where each subplot represents the search results for one energy interval data set. The top two panels of each subplot represent the power spectra obtained by the SFFT algorithm and the TDA when $f_1$ is not considered, respectively, and the heatmap below represents the results of the search for $f_0-f_1$ using the TDA. The bright line in the heatmap indicates that a peak in the power spectrum drifts in the frequency space as the searched $\frac{f_1}{f_0}$ changes. The profiles were calculated using the best $f_0$ and best $f_1$ for each data set, and none of the profiles exceeded the 3 $\sigma$ significance level (assuming a Poisson distribution). This means that there are no significant periodic signals at 0.0017 Hz$\sim$ 70 Hz in this data set.

\subsection{Long Timescale Period Search}
The Lomb-Scargle (LS) algorithm is used to search for periods of the non-uniformly sampled signals \cite{Lombscargle_1976ApSS..39..447L,Lombscargle_1982ApJ...263..835S}.
For long-term period search, the discontinuity of observation will inevitably lead to the non-uniform sampling. For the ordinary FFT algorithm, the gap can be filled by filling the zeros, but it is more appropriate to use the LS algorithm for the nature of the low burst rates. We use the LS algorithm to search for long periods of ToA series and to ensure the robustness of the results, the results are confirmed by a double check of the TDA processing.

For the long timescale periodicity search, given that the entire observation period is approximately 216 days, to ensure that there are still at least 2 periods in the data even in the case of the longest period, we set the maximum period of the search to 108 days (resulting in a minimum search frequency of about 0.0093 $\rm days^{-1}$). As for the maximum search frequency, we set it rather arbitrarily. Taking into account the shortest duration of the continuous observations, which is 1200 seconds, we set the time bin width to 0.01387 days, leading to a maximum search frequency of 36 $\rm days^{-1}$.

\begin{figure*}
    \centering
    \includegraphics[scale=0.45]{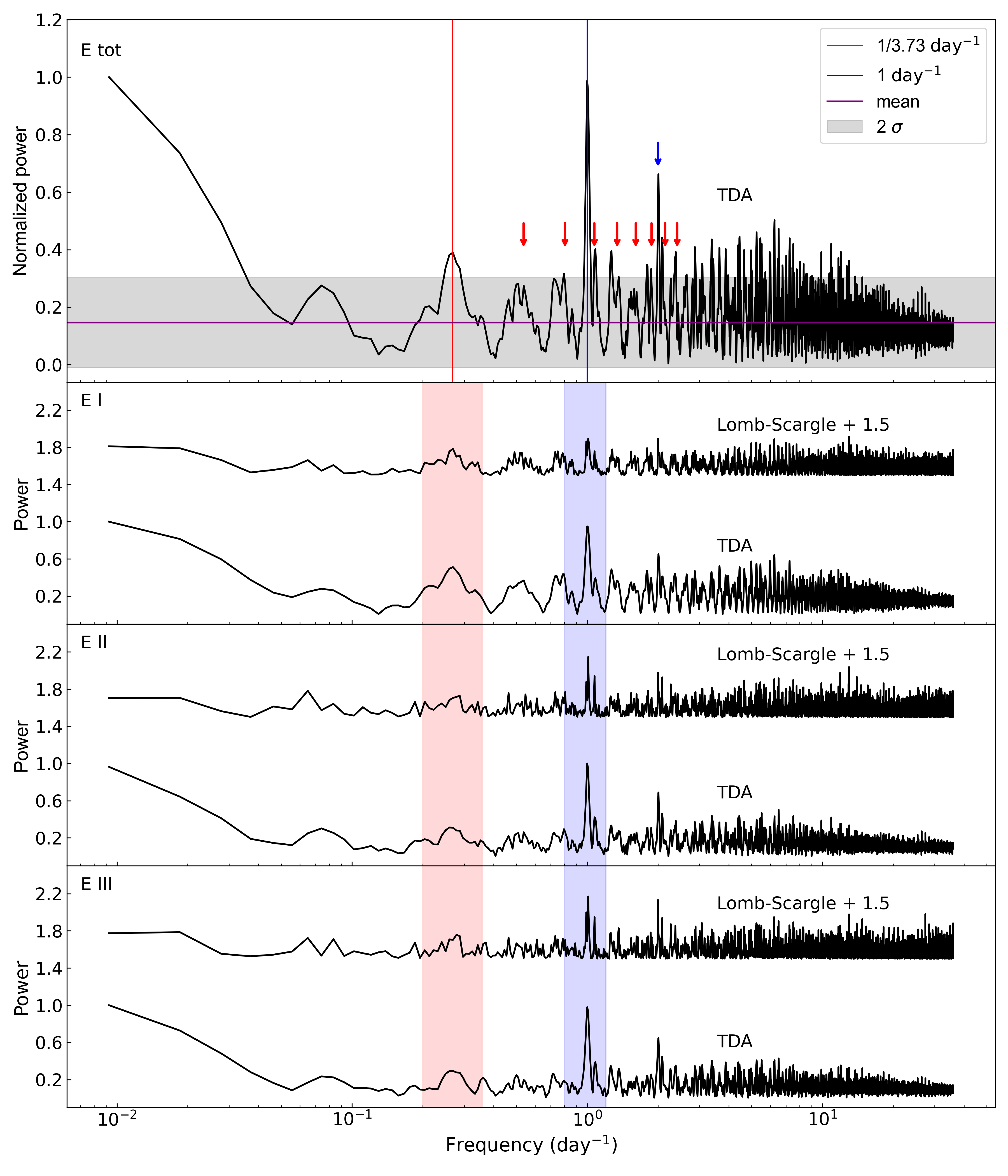}
    \caption{{\bf The result of the long timescale periodicity search using the Lomb-Scargle and the TDA.} Each sub-panel corresponds to a dataset from different energy intervals. The top panel shows the power spectrum of the TDA search for all of the 555 bursts (``E tot"), and the following sub-panels represent the search results for different energy regions for ``E I", ``E II" and ``E III" from top to bottom. The red and blue lines indicate frequencies of 1/3.7 ${\rm day^{-1}}$ and 1 $\rm day^{-1}$, respectively, and the corresponding harmonics are also marked with arrows. In the sub-panels for each energy region, the upper and lower curves separately depict the power spectra using the Lomb-Scargle and the TDA. The shades of red and blue correspond to the frequencies of 1/3.7 ${\rm day^{-1}}$ and 1 $\rm day^{-1}$, respectively.}
    \label{EDfig4}
\end{figure*}

The results of the long timescale periodicity search using the TDA and the Lomb-Scargle algorithms are shown in Extended Data Figure \ref{EDfig4}. The top panel shows the power spectrum of the TDA search for all of the 555 bursts (``E tot"), and the following sub-panels present the TDA and Lomb-Scargle search results for energy regions in ``E I", ``E II" and ``E III" from top to bottom. It can be seen that both algorithms give apparent peaks at 1 $\rm day^{-1}$ and its harmonic frequencies (marked with blue lines, blue arrows, and blue shade). The peaks are attributed to the cadence of the FAST observations, i.e., observations were taken at roughly the same time for each day (e.g., from MJD 59834 to MJD 59840), resulting in a spurious 1-day periodicity. On the other hand, a low-significance quasi-periodic peak of around 1/3.7 $\rm day^{-1}$ can also be seen at the lower frequencies of the power spectrum (marked with red lines, red arrows, and red shade), which is also attributed to the observing cadence. Besides these spurious periods, there are no other reliable periodic signals in the power spectrum, indicating that the repeating FRB 20121102A does not have a credible periodicity on the long timescale of 0.028 days $\sim$ 108 days during the FAST observing campaigns in 2022 and 2023.

\section{A Comprehensive Analysis}
We present a detailed comprehensive analysis and the potential correlation among various measured parameters of the repeating FRB 20121102A (see Supplementary Table 2 for the full catalog) in this section. $\sim$65$\%$ of the bursts are typically detected over less than half of the 500 MHz observation bandwidth, suggesting that narrow-band bursts may be more common than in previous studies of FRB 20121102A \cite{gourdji19, petroff22, Oostrum20, cruces21, Caleb19, Chamma23, Hewitt22, Kshitij21}. Extended Data Figure \ref{EDfig5} indicates the burst spectrum full width at half maximum (FWHM) as a function of burst number and its histograms in the top panel. The blue and yellow histograms represent the burst detection from the observations of Episode I (2022) and Episode II (2023), respectively. We show that the spectral characteristics of the bursts do not evolve significantly in time over the two episodes, and note that the bursts occur at preferred frequencies in 1.25-1.5 GHz. The narrowest frequency bandwidth of all 555 pulses detected is 70 MHz, with a median of 250 MHz.

\begin{figure}
     \centering
    \includegraphics[scale=0.65]{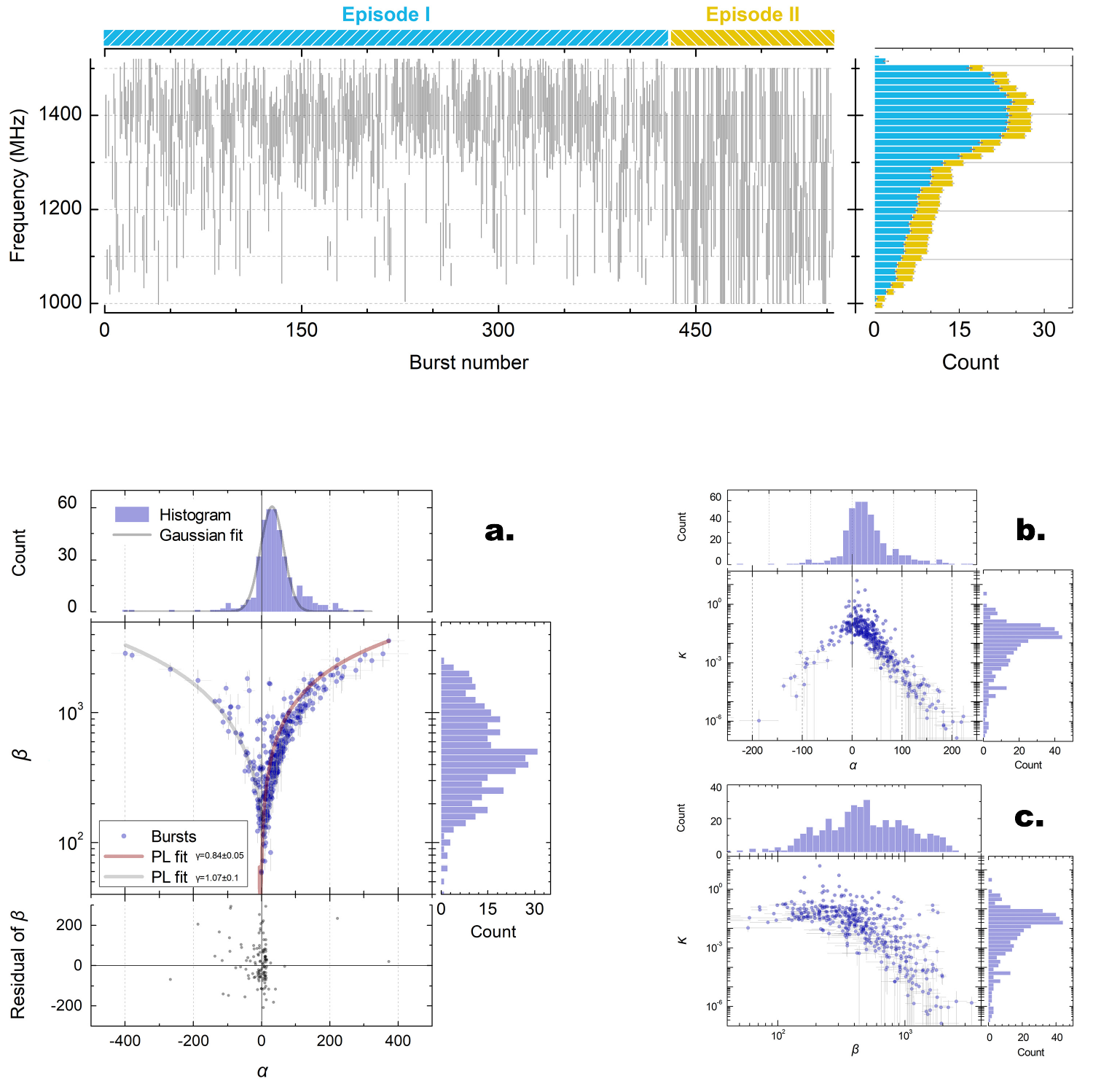}
    \caption{{\bf The top panel shows the burst spectrum (FWHM) as a function of burst number and its histograms.} The blue and yellow histograms represent the burst detection from the observations of Episode I (2022) and Episode II (2023), respectively. Error bars in histograms are derived using the bootstrap method. {\bf The bottom panel indicates the pulse morphology and the spectrum fitting parameters.} Bursts are plotted separately on different parameter planes, and the error bars are calculated at the 68$\%$ confidence level. Panel a: a clear separation of the narrow- and broad-band populations suggests $\beta$ is a discriminating parameter between them. Two-component power-law (PL) distribution is separately fitted in red and grey solid lines, the fitting residual is at the bottom of the panel a.}
    \label{EDfig5}
\end{figure}

We quantify the burst spectrum with a signal to noise ratio (S/N) $\textgreater$10 using the phenomenological function $I = \kappa(\frac{\nu}{\nu_0})^{\alpha+\beta \log(\frac{\nu}{\nu_0})}$, where $\nu$ is a frequency value between 1.0 and 1.5 GHz of FAST L-band receiver, $I$ is the amplitude (arbitrary units) at a given frequency, $\alpha$, $\beta$, and $\kappa$ are three of the free parameters and $\nu_0$ = 1.25 GHz is the reference frequency. 
The fitting function would degenerate to a power-law function with $\beta$$\sim$$0$, a higher $\beta$ indicates narrower band emission.
We estimate the fitting uncertainties through the standard error propagation with MCMC fitting to avoid overfitting. 
The distributions of the morphological fitting parameters $\alpha$ and $\beta$ are shown in the bottom panel of Extended Data Figure \ref{EDfig5}. 
The distribution of $\beta$ against $\alpha$ separates the narrow-band and broad-band pulses (Extended Data Figure \ref{EDfig5}a). 
The narrow-band pulses (higher $\beta$) tend to have a steeper spectrum (higher $\mid$$\alpha$$\mid$) than the broad-band ones, and the fraction of the bursts with a positive spectral index is higher than 85$\%$.
A power-law (PL) distribution ($\beta$=A$\mid$$\alpha$$\mid^{\gamma}$) is fitted to the two branches ($\alpha$$\leq$0 and $\alpha$$\geq$0) with respective indexes of $\gamma$=0.84$\pm$0.05 and 1.07$\pm$0.1, and the residuals have no obvious structure. 
Furthermore, the distribution of $\kappa$ against $\beta$ and $\alpha$ suggests these narrow-band bursts are mainly concentrated at low flux densities and with multiple components in the panel b and c of Extended Data Figure \ref{EDfig5}, which corresponds to the same trend in other repeating FRBs \cite{pleunis21}.

We investigate the potential correlation among various key parameter pairs of the bursts, which include the DM, center frequency, bandwidth, pulse width, peak flux density, and isotropic equivalent energy. By exploring the statistical distribution of the double parameters, we can place further constraints on the triggering and the emission mechanism of repeating FRBs. 

Extended Data Figure \ref{EDfig6}a shows the pair-wise correlation analysis between the measured parameters of the FRB 20121102A. The blue, yellow, and grey histograms represent the FAST detection from the 2022, and 2023 observational campaigns, and all the detected 555 bursts, respectively. The red lines in each panel indicate the estimated probability density contours using the Kernel Density Estimation (KDE) method, while the blue dots represent the observed bursts. There is no obvious correlation between almost all the parameter pairs, which means that most of the parameters are incoherent with each other, except in the peak flux density - isotropic equivalent energy and the pulse width - isotropic equivalent energy planes. 

We calculate the Pearson correlation coefficient (PCC) for each of the double parameter pairs, which shows that the PCC values are higher than 0.5 only for the flux density - energy plane and the pulse width - energy plane, namely 0.51 and 0.54, respectively. The positive correlations can be attributed to the fact that the isotropic equivalent energy is proportional to the fluence, which is equal to the product of the peak flux density and the effective pulse width $W_{eff}$. The flux density distribution spans a small range of magnitudes on the histogram (within an order of magnitude), while the $W_{eff}$ distribution spans nearly 3 magnitudes, resulting in the $W_{eff}$ distribution dominating the energy distribution. On the other hand, there is no positive correlation between the peak flux density and $W_{eff}$, which is inconsistent with the neutron star-asteroid collision models \cite{dai20} of repeating FRBs.

Comparing the distributions of the histograms for each parameter in episodes the 2022 and 2023, we note that the differences in the distributions of the parameters, especially the histogram peak locations, are almost negligible, except for the DM. The DM histogram shows an observable bimodal distribution in two of the adjacent episodes, i.e. the peak location evolves from DM = 553 $pc\ cm^{-3}$ (Episode I, 2022) to 552.6 $pc\ cm^{-3}$ (Episode II, 2023) over a period of about 110 days, suggesting a systematic and continuous decreasing tendency of DM. We also note that there is no correlation between the DM and other parameters measured from the bursts, such as energy, flux density, time-frequency morphological features, etc. Consequently, the temporal evolution of the DM tends to reflect dynamic variations in the local environment rather than being due to the central engine of the repeating FRB 20121102A.

B489 is the only burst with RM measured in the FAST observing campaign, and a total of 45 bursts (B482-B526) were detected during the 20-minute duration of observation on the same day (3rd Mar. 2023). In the corner plots of Extended Data Figure \ref{EDfig6}b, the flux density and isotropic equivalent energy of B489 is more than two orders of magnitude higher than the other bursts of that day. Under the assumption that FRB has multi-path propagation-induced depolarization \cite{Feng22}, the Stokes I of the bursts need to be greater than 0.44 Jy for a 7$\sigma$ detection on the Faraday spectrum for a linear polarization component like that of the B489. The non-detection of the linear polarization can be attributed to the low flux density of the other bursts detected on 3rd Mar. 2023. We further checked the bursts with comparable flux densities to B489 at adjacent times: B464 (February 20, 2023) and B536 (March 20, 2023), and unfortunately no polarization was detected in both bursts. This implies that we are hard to make constraints on burst-to-burst or day-to-day variations on the propagation effect of the magnetized plasma, and can only give a conservative upper limit on the temporal evolution timescale of about two weeks.

\begin{figure}
     \centering
    \includegraphics[scale=0.45]{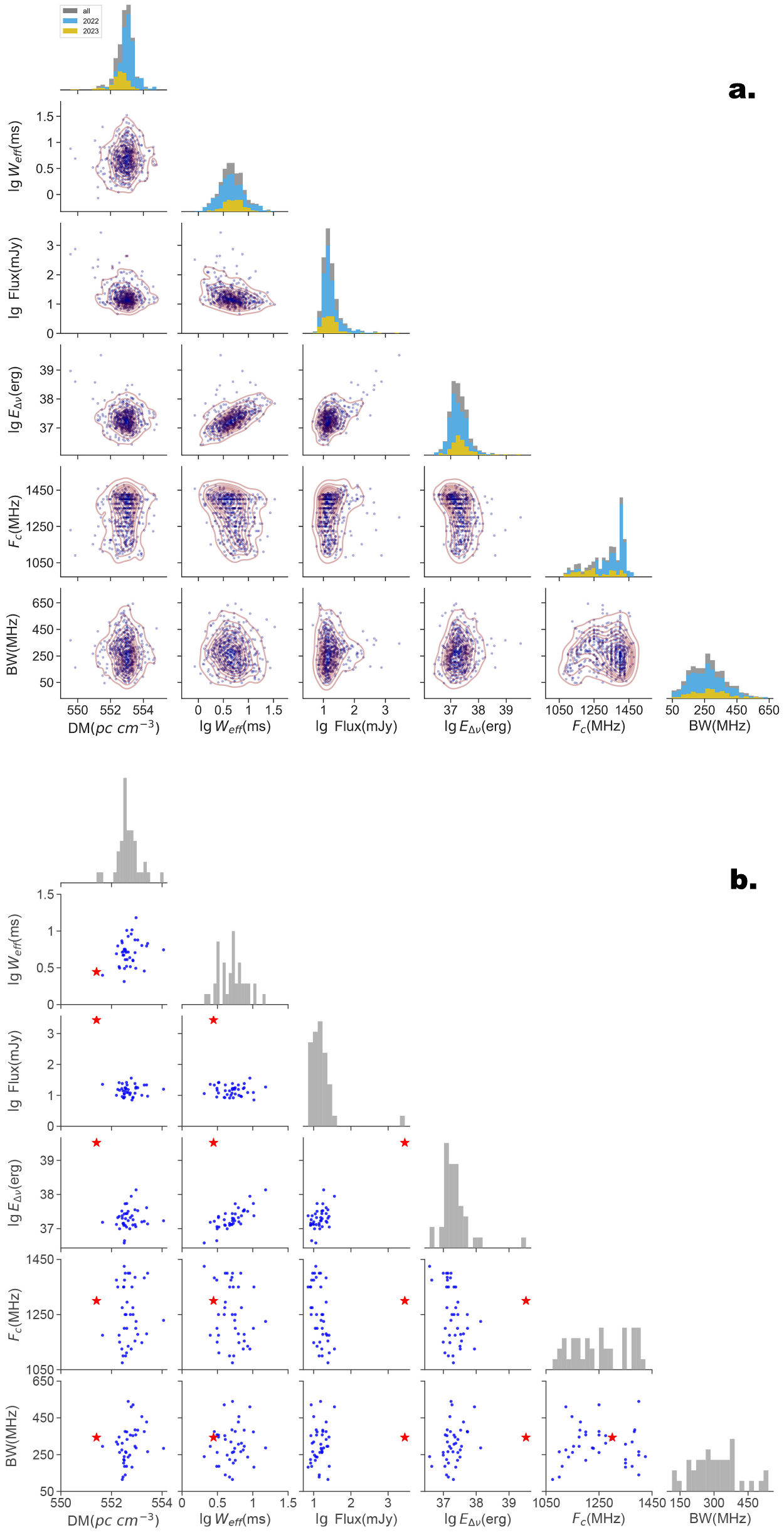}
    \caption{{\bf The pair-wise correlation analysis between the measured parameters of FRB 20121102A.} Panel a: the blue, yellow, and grey histograms represent the data set from the 2022, and 2023 observation episodes, and all the 555 bursts, respectively. The red contour lines present in each panel are the result of the KDE fitting. Panel b: same as Panel a, except that the data is replaced with 45 bursts detected on the 3rd Mar. 2023. The red star is the burst of B489.}
    \label{EDfig6}
\end{figure}

\section{Morphology}
In this section, we present an analysis of the burst morphology, i.e. the change of the flux density and energy as a function of time and frequency, as detected with the FAST L-band (1.0-1.5 GHz) during the two active episodes in 2022 and 2023, using a total of 555 bursts from the repeating FRB 20121102A (see Supplementary Table 2 for the full catalogue). Using the catalogue properties of the bursts, we apply a novel statistical clustering approach to obtain a more quantitative understanding of the burst morphology and the possible implications of these morphological differences.

The Bisecting k-means (bi-Kmeans) algorithm is a clustering algorithm that recursively splits the dataset into smaller clusters using k-means clustering.
The algorithm starts with one cluster and splits it into two until a desired number of clusters is reached. 
We use the bi-Kmeans approach to cluster the multiple key parameters of the bursts, including DM, center frequency, bandwidth, pulse width, peak flux density, and isotropic equivalent energy. 
Since the bi-Kmeans requires the number of clusters to be specified, a metric is needed to evaluate the goodness of the results for different numbers of clusters. 
The Calinski-Harabasz (CH) index is one of the most common evaluation metrics used in clustering algorithms to measure the quality of clustering results. 
It measures the balance between the tightness within clusters and the separation between different clusters. The CH index for K number of clusters on a dataset D =[ d$_1$, d$_2$, d$_3$, … d$_N$] is defined as,
\begin{equation}
    CH = \frac{(N-K)\sum_{k=1}^Kn_k\Vert
        c_k-c\Vert^2}{(K-1)\sum_{k=1}^K\sum_{i=1}^{n_k}\Vert d_i-c_k\Vert^2},
\end{equation}
where, $n_k$  and $c_k$ are the number of the point and centroid of the $k$-th cluster respectively, $c$ is the global centroid, $N$ is the total number of data points.

A high value of the CH index means that the clusters are dense and well separated. 
For different numbers of clusters, we use the bi-Kmeans algorithm to cluster the data, calculate the corresponding CH index, and finally get the CH index as a function of the number of clusters, as shown in Extended Data Figure \ref{EDfig7}a. 
We find that the CH index reaches a maximum value when the number of clusters is 4, which means that it is the optimized clustering of the dataset into 4 classes.

\begin{figure}
    \centering
    \includegraphics[scale=0.16]{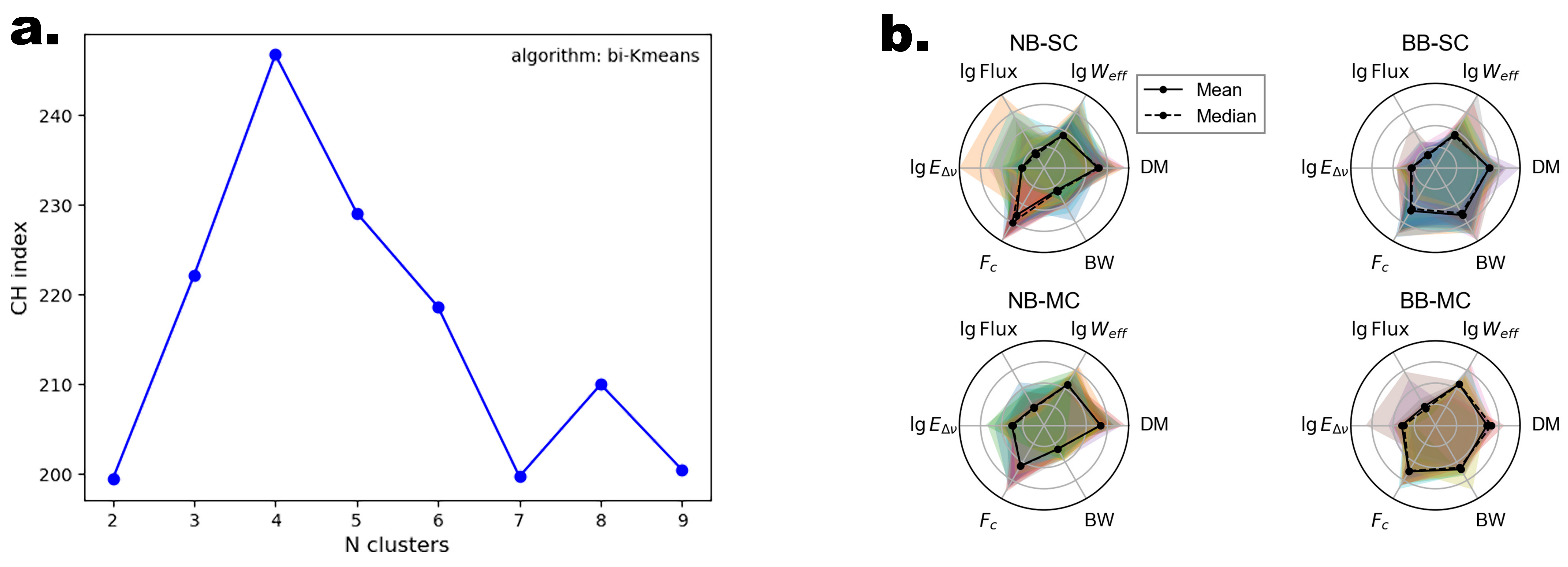}
    \caption{{\bf Result of morphological clustering using the bi-Kmeans algorithm.} Panel a: the CH index as a function of the number of clusters; Panel b: The radar charts of the distribution of key observed parameters, classified into 4 morphologies. The parameter sets corresponding to the bursts are marked in different colors in each panel, and the black solid and dashed lines represent the mean and the median values of each parameter in each clustered panel.}
    \label{EDfig7}
\end{figure}

Specifically, all the bursts can be classified into 4 morphology types: Broad-Band (BB), Narrow-Band (NB), Single-Components(SC), and Multiple-Components (MC), according to the results of the bi-Kmeans algorithm. 
In Extended Data Figure \ref{EDfig7}b, we use the radar charts to visually compare the difference in burst parameters for different morphology types.
Each colored polygon represents the parameters of a specific burst, and the solid black line represents the average of each parameter, while the dashed black line represents the median value. 
We normalize each observed parameter separately on the radial scale of Extended Data Figure \ref{EDfig7}b by the maximum value of all the FAST detected bursts from 2022 to 2023. 
The implementation of normalized parameters allows a quantitative comparison of the parameters for morphological types in radial coordinates.
We find that the NB-SC bursts tend to occupy lower frequencies compared to the NB-MC bursts, while the opposite is seen for BB bursts. 
Furthermore, SC bursts typically have lower flux densities and energies compared to MC bursts.

To further reveal the possible implications of the morphological differences, we present the distribution of the bursts in the $BW/F_c - \log(E\ W_{eff})$ plane, and use the kernel density estimation (KDE) approach to estimate the probability density distribution of each morphological type in the left panel of Extended Data Figure \ref{EDfig8}. 
The colored contour-shaded areas and dots indicate the probability density distribution and the centroid for each morphological type respectively, the red cross mark is the global centroid. 
The shaded areas, although still partially overlapping, show a clear distinction in the occupation of the $BW/F_c - \log(E\ W_{eff})$ plane.
In the right panel of Extended Data Figure \ref{EDfig8}, we can further quantify the probability density function of each morph type by polar coordinate transformation.
We define the global centroid as the center of the polar coordinate and define the radial distance $r$ and the angle $\theta$ as
\begin{gather}
    \theta = \arctan(\frac{\lg E_{\Delta \nu }W_{eff} - \lg E_{0}W_{0}}{10BW/F_{c} - 10BW_0/F_{0}}),\\
    r = \sqrt{(10BW/F_{c} - 10BW_0/F_{0})^2 + (\lg E_{\Delta \nu }W_{eff} - \lg E_{0}W_{0})^2}.
\end{gather}
The grey arrows indicate the vectors from the global centroid to the centroid of each type, and the angle $\theta$ between the arrows indicates the distinction between morphological types. The probability density function $P(r,\theta)$ for each morph type $P_i$ has 
\begin{equation}
    P_i = \frac{\int^{\theta_{ui}}_{\theta_{li} }\int^{l_i}_0 P(r,\theta) dr d\theta }{ \int_0^{2\pi} \int_0^\infty P(r,\theta) dr d\theta} ,
\end{equation}  
where $l$ is the upper limit of the radius, $\theta_l$ and $\theta_u$ denote the lower and the upper limit of the angle, respectively.

Combined with Figure \ref{fig1}, neither the isotropic equivalent energy (panel c) nor the $BW/F_c$ (panel d) show any notable time dependence during the observation episodes, making it possible that the observed differences of burst morphology are intrinsic, with the morphological types forming four distinct populations of bursts produced by different emission mechanisms or short time-scaled propagation effects, e.g. Cordes et al. \cite{cordes16} proposed a plasma lensing model to explain the complex time-frequency patterns.

\begin{figure}
    \centering
    \includegraphics[scale=0.6]{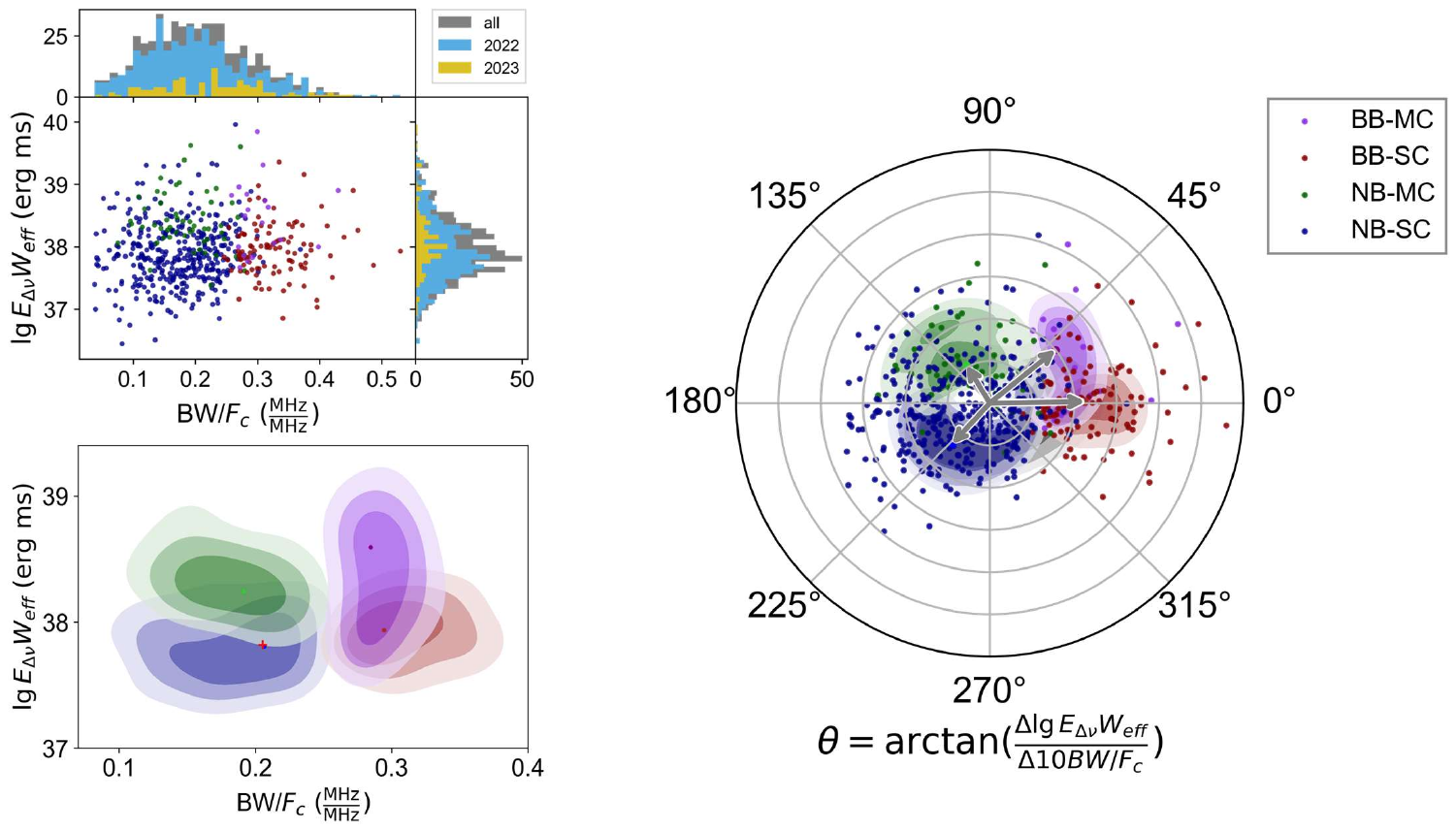}
    \caption{{\bf The distribution of the typical morphology types.} Left panel: the distribution of the typical morphology types in the $BW/F_c - \log(EW_{eff})$ plane. The colored contour-shaded areas and dots indicate the probability density distribution and the centroid for each morphological type respectively, the red cross mark is the global centroid. Right panel: the distribution of the probability density function of the morphological types in polar coordinates.}
    \label{EDfig8}
\end{figure}

 \section{Polarization Analysis}
The polarization of the detected bursts was calibrated by correcting for the differential gains and phases by separate measurements of a noise diode injected at an angle of $45^{\circ}$ from the linear receiver. The circular polarization is consistent with noise, lower than a few percent of the total intensity, which agrees with Ref.\cite{michilli18}.
We perform a search for the RM within a range of $-5.0\times10^5$ to $5.0\times10^5$\,$\mathrm{rad\,m^{-2}}$, which safely covers the value of $\mathrm{RM} \sim 10^5\,\mathrm{rad\,m^{-2}}$ reported in Ref.\cite{michilli18}.

\begin{figure*}
    \centering
    \includegraphics[scale=0.65]{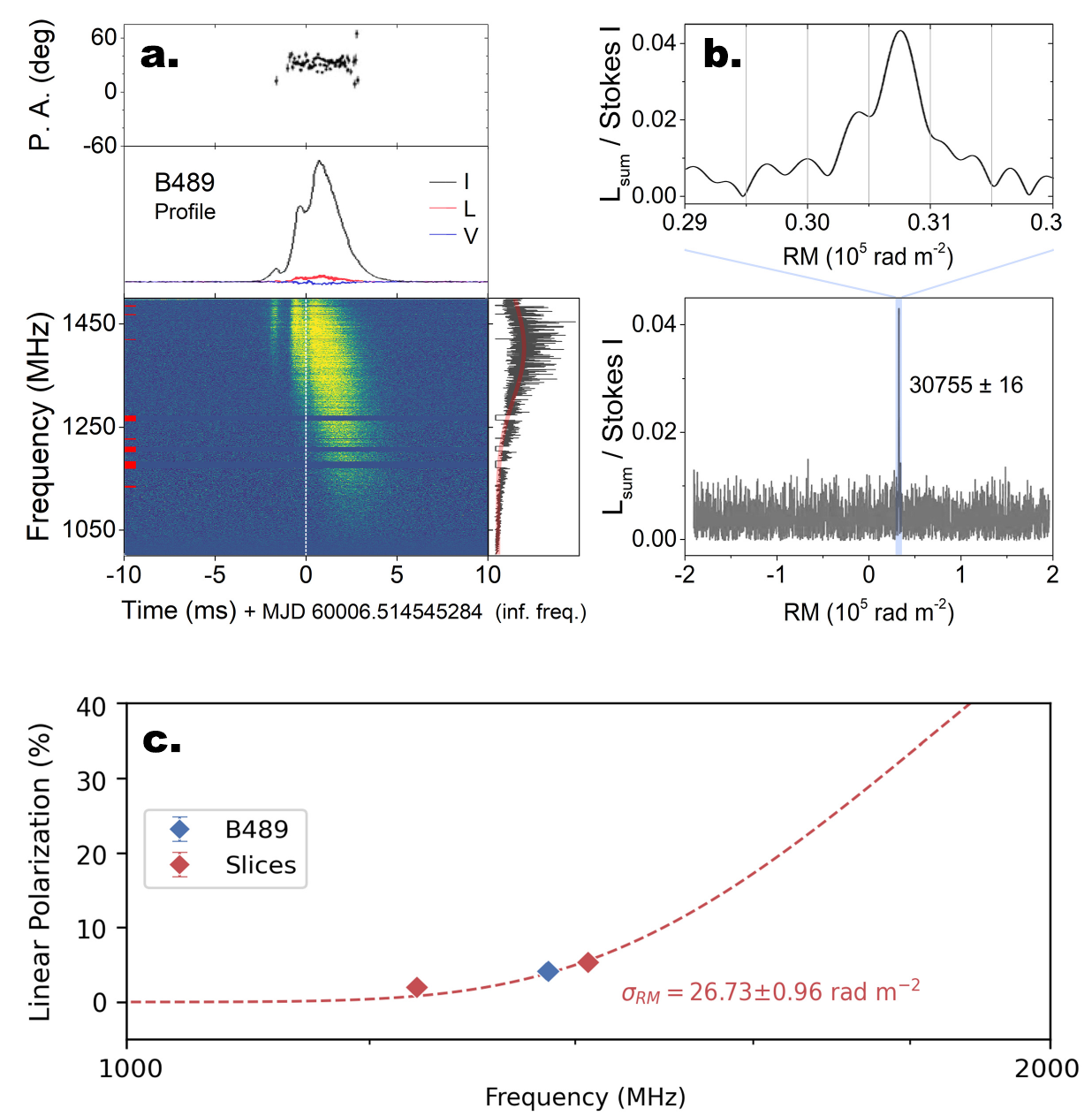}
    \caption{{\bf Polarization detection of B489 using the FAST L-band.} Panel a: the sub-panels from top to bottom show the polarization position angle (PA) and the polarization pulse profiles, with dynamic spectra for the burst of B489 detected by FAST (1-1.5 GHz). The black/red/blue lines separately demonstrate the profiles of total intensity, linear polarization, and circular polarization in the middle sub-panel. Panel b: RM fit and a magnified plot for B489 is shown by the linear polarization percentage as a function of the RM value. Panel c: the $\sigma_{RM}$ of B489 was fitted using frequency slicing.}
    \label{EDfig9}
\end{figure*}

Except for the burst of B489, no significant peak was found on the Faraday spectrum. 
B489 has the highest isotropic equivalent energy of all the 555 bursts detected in the 2022-2023 FAST observing campaign. The linear polarization fraction of B489 is about 4\% at RM = 30755 $\mathrm{rad\,m^{-2}}$ with an error of 16 $\mathrm{rad\,m^{-2}}$ in Extended Data Figure \ref{EDfig9}, and indicating that the search for linear polarized component of the bursts should be limited to those with high flux densities.
The linear polarization becomes negligible in the L-band compared to almost 100\% linear polarization in the C-band reported in Ref.\cite{michilli18}. 
In this work, we estimate the depolarization fraction $f_{\rm depol}$ using 
\begin{equation}
\label{eq:depol}
f_{\rm depol} = 1 - \frac{{\rm sin}(\Delta{\theta})}{\Delta{\theta}},
\end{equation}
where the intra-channel Faraday rotation $\Delta{\theta}$ is given by

\begin{equation}
\label{eq:faraday_smearing}
\Delta\theta = \frac{2{\rm RM}_{\rm obs}c^2\Delta\nu}{\nu_c^{3}},
\end{equation}
where $c$ is the speed of light, $\Delta{\nu}$ is the channel width, and $\nu_c$ is the central channel observing frequency. Taking $\Delta{\nu} = 0.122\,\mathrm{MHz}$, and $\nu_c = 1.25\,\mathrm{GHz}$ for our data, we get $f_{\rm depol}$ = 20$\%$ and 6$\%$ for ${\rm RM}_{\rm obs}$ = 10$^5$ rad m$^{-2}$ as reported in Ref.\cite{michilli18}, and 30755 rad m$^{-2}$. Although the depolarization fraction is not negligible, the non-detection of the linear polarization cannot be caused by depolarization assuming ${\rm RM}_{\rm obs} \leq$ 10$^5$ rad m$^{-2}$.

We are confident about the non-detection of polarization in most of the bursts in this work. We have performed the same analysis procedures to the multiple bright repeating FRBs and pulsars, and retrieved expected results \cite{Feng22,luo20nat,zhangCF20,xu22,jiang23}. 
The depolarization is quantified by the parameter $\sigma_{RM}$ \cite{Feng22}, which is related to the propagation process in the magnetized, inhomogeneous plasma medium around the FRB. The 4\% linear polarization ratio of B489 and the non-detection of other bursts do not conflict with the previous almost 100\% linear polarization, but indicate a strong frequency evolution of the linear polarization: the degree of the linear polarization for FRB 20121102A is consistent with depolarization due to the RM scattering \cite{Feng22}, and all previous polarization detections were accomplished at frequency bands higher than L-band ($\sim$1.4\, GHz). Ref.\cite{michilli18} published the polarization measurements at 4-8\,GHz and Ref.\cite{hi21} at 3-8\,GHz. 
The value of $\sigma_{RM}$ = 26.73$\pm$0.96 rad m$^{-2}$ (panel c of the Extended Data Figure \ref{EDfig9}) is shown to have decreased by $\sim$13$\%$ during the 2022-2023 FAST observing campaigns as compared to previous determinations of $\sigma_{RM}$ = 30.9 rad m$^{-2}$, thus reflecting the evolution of the inhomogeneous magneto-ionic environment. 
The bursts with comparable flux densities and adjacent times to B489 (3rd Mar. 2023) are B464 (20th Feb. 2023) and B536 (20th Mar. 2023), unfortunately, no polarization has been detected in either of those bursts. This also implies that the inhomogeneous magneto-ionic environment reflected in the value of $\sigma_{RM}$ is time-sensitive, with an upper limit on the temporal evolution timescale of about two weeks.

\section{Physical Origin of the Evolution of DM and RM of FRB 20121102A}

\subsection{Observation of FRB 20121102A}\label{sec_observation}

FRB 20121102A has an observed DM of ${\rm DM_{obs}}\gtrsim 552~{\rm pc~cm^{-3}}$ and a DM variation of $d{\rm DM_{obs}}/dt=(-3.93-0.86)~{\rm pc~cm^{-3}yr^{-1}}$ from 2022 Aug. to 2023 Mar. The long-term DM evolution from 2012 to 2023 is $d{\rm DM_{obs}}/dt=-0.43~{\rm pc~cm^{-3}yr^{-1}}$, which is directly measured by the DM of ${\rm DM_{obs}}\simeq557~{\rm pc~cm^{-3}}$ in 2012 \cite{Spitler2014} and the latest measurement in this work.
Based on the host-galaxy redshift of $z=0.193$ \cite{tendulkar17}, the DM contribution of the interstellar medium (IGM) of FRB 20121102A is about ${\rm DM_{IGM}}\simeq164~{\rm pc~cm^{-3}}$ (adopting the Planck cosmological parameters and $f_{\rm IGM}=0.83$, see \cite{yangzhang17} for details) and the Milky Way contribution is about ${\rm DM_{MW}}=218~{\rm pc~cm^{-3}}$ \cite{chatterjee17}. Therefore, making the redshift correction of the host contribution ${\rm DM_{host}}$, the local DM contributed by the FRB environment satisfies
\be
{\rm DM}\lesssim{\rm DM_{host}}={\rm DM_{obs}}-{\rm DM_{MW}}-{\rm DM_{IGM}}\simeq203~{\rm pc~cm^{-3}}.
\ee
We should notice that the above value still has a relatively large uncertainty from the IGM fluctuation, $\sigma_{\rm IGM}\sim0.2{\rm DM_{IGM}}z^{-1/2}$ \cite{Kumar19}. 
Since the DM variations contributed by the IGM and the ISM are extremely small as proposed \cite{yangzhang17}, the observed DM evolution could only be caused by the local plasma by the FRB environment, leading to
\be
\frac{d{\rm DM}}{dt}\simeq\frac{d{\rm DM_{obs}}}{dt}=(-3.93-0.86)~{\rm pc~cm^{-3}yr^{-1}}.\label{DMobs}
\ee
On the other hand, the observed extremely large RM and the significant RM evolution could only be caused by the local plasma by the FRB environment \cite{yang23}, and the observation shows 
\be
{\rm RM}\sim10^{5}~{\rm rad~m^{-2}},~~~\frac{d{\rm RM}}{dt}\simeq-10^4~{\rm rad~m^{-2}yr^{-1}}.\label{RMobs}
\ee
Based on Eq.(\ref{DMobs}) and Eq.(\ref{RMobs}), one finally has
\be
\left|\frac{d\ln{\rm DM}}{dt}\right|\gtrsim(4-19)\times10^{-3}~{\rm yr^{-1}},~~~\left|\frac{d\ln{\rm RM}}{dt}\right|\simeq0.1~{\rm yr^{-1}}.
\ee
The values of $|d\ln{\rm DM}/dt|^{-1}$ and $|d\ln{\rm RM}/dt|^{-1}$ correspond to the typical timescales of the DM and the RM evolutions, respectively.

\subsection{Evolution of DM and RM from a Young Supernova Remnant}\label{sec_snr}

FRB 20121102A has the largest $|{\rm RM}|$ among all observed FRB sources, and the observed RM decreased by $|\delta{\rm RM}|\sim10^5~{\rm rad~m^{-2}}$ within a few years with the RM sign remained \cite{michilli18,hi21}. Such a monotonic RM evolution could be naturally attributed to an expanding supernova remnant (SNR) with unchanged field configuration \cite{piro18,hi21,yang23}. During the SNR expansion, both the electron density and the parallel magnetic field would evolve, finally leading to a power-law evolution of the RM,
\be
{\rm RM}\propto t^{-\alpha},
\ee 
where the temporal index $\alpha$ depends on both the states of the gas and the magnetic field and the evolution stage of the SNR \cite{piro18,yang23}. According to the above power-law evolution, the SNR age could be obtained by
\be
t_{\rm SNR}=\alpha\left|\frac{dt}{d\ln{\rm RM}}\right|.\label{tRM}
\ee
The observation of FRB 20121102A shows $|d\ln{\rm RM}/dt|\simeq0.1~{\rm yr^{-1}}$, giving an estimate of  the SNR age of $t_{\rm SNR}\sim10\alpha~{\rm yr}$. 
Since the temporal index $\alpha$ is of the order of unity \cite{piro18,yang23}, the fast monotonic RM evolution of FRB 20121102A implies that the associated SNR is young with an age of a few $\times(1-10)$ years, consistent with the measured age of the source $\gtrsim 12$ years.
Such a young SNR must be in the free-expansion stage, and both the DM and RM are dominated by the SN ejecta rather than the shocked interstellar medium. 

Two possible scenarios were analyzed for the young SNR in the free-expansion phase in the literature: fully ionized SN ejecta \cite{Katz16, Murase16, yangzhang17, yang23} and shocked ionized SN ejecta \cite{piro18}. We consider that the SN ejecta has a mass of $M$, a kinetic energy of $E$, a mean molecular weight per electron of $\mu_e=1.2$ for the solar composition, and an initial magnetic field of $B_0$. The ejecta velocity is $v=(2E/M)^{1/2}$. In the following discussion, the convention $Q_x = Q/10^x$ is adopted in cgs units unless otherwise specified.

I) Fully ionized ejecta: If the SN ejecta is fully ionized in the free-expansion phase, all electrons in the SN ejecta would contribute to the DM and RM. A time-dependent DM contributed by a young SNR is
\be
{\rm DM}=\frac{M}{4\pi\mu_e m_p(vt)^2}\simeq260~{\rm pc~cm^{-3}}M_{0,\odot}^2E_{51}^{-1}t_{10{\rm yr}}^{-2},
\ee
where $M_{0,\odot}=M/1M_\odot$ and $t_{10{\rm yr}}=t/10{\rm yr}$, and the DM variation of the SNR is
\be
\frac{d{\rm DM}}{dt}=-\frac{M}{2\pi\mu_e m_pv^2t^3}\simeq-52~{\rm pc~cm^{-3}yr^{-1}}M_{0,\odot}^2E_{51}^{-1}t_{10{\rm yr}}^{-3}.
\ee
The associated RM is
\be
{\rm RM}=\frac{e^3}{2\pi m_e^2c^4}\int n_eB_\parallel dl\simeq0.81~{\rm rad~m^{-2}}B_{\parallel, -6}{\rm DM},
\ee
where ${\rm DM}$ is in the unit of ${\rm pc~cm^{-3}}$. For a fully ionized SN ejecta, the evolution of the magnetic field in the SN ejecta is complex. For simplicity, we discuss two possible cases here: the magnetic field frozen in the SN ejecta and the magnetic field satisfying energy-equipartition with the ejecta gas. 1) For the magnetic freezing scenario, the magnetic field satisfies
\be
B_\parallel(r)=\xi B(r)=\xi B_*\fraction{r}{R_*}{-1},
\ee
where $B_*$ is the toroidal magnetic field in the progenitor envelope at radius $R_*$, and $\xi\equiv \left<B_\parallel\right>/\left<B\right>$ is a parameter depending on the field geometry. We notice that the toroidal magnetic field is dominant because of its relatively slow decay in the free-expansion phase, thus, the contribution from the radial field could be ignored here. Thus, the corresponding RM is
\be
{\rm RM}\simeq\frac{e^3}{2\pi m_e^2c^4}\frac{M\xi B_*R_*}{4\pi\mu_e m_p(vt)^3}\simeq46~{\rm rad~m^{-2}}\xi M_{0,\odot}^{5/2}E_{51}^{-3/2}B_{*,0}R_{*,\odot}t_{10{\rm yr}}^{-3},
\ee
where $R_{*,\odot}=R_*/R_\odot$,
and the RM variation of the SNR is
\be
\frac{d{\rm RM}}{dt}\simeq-\frac{e^3}{2\pi m_e^2c^4}\frac{3M\xi B_*R_*}{4\pi\mu_e m_pv^3t^4}\simeq-14~{\rm rad~m^{-2}yr^{-1}}\xi M_{0,\odot}^{5/2}E_{51}^{-3/2}B_{*,0}R_{*,\odot}t_{10{\rm yr}}^{-4}.
\ee
We can see that the RM contributed by the frozen magnetic field in the SN ejecta is relatively small. We notice that for a toroidal field, there might be a vertical component because of the dissipation of the toroidal field. However, the vertical component might be a small fraction, leading to $\xi\ll1$ and a smaller RM value.

2) For the magnetic field satisfying energy-equipartition with the ejecta gas, one has $B^2/8\pi\sim\epsilon_BE/V$ with $V\sim(4\pi/3)r^3$, where $\epsilon_B$ is a parameter that sets how much of the shock energy goes to the magnetic field, leading to
\be
B_\parallel\sim\xi B\fraction{6\epsilon_BE}{r^3}{1/2}.
\ee
The scenario requires the operation of a mechanism capable of reaching energy equipartition between magnetic fields and particles on a timescale short compared with the dynamical and radiative timescales.
The corresponding RM is
\be
{\rm RM}\simeq\frac{e^3}{2\pi m_e^2c^4}\frac{3^{1/2}\xi\epsilon_B^{1/2}M^{3/2}}{4\pi\mu_e m_pv^{5/2}t^{7/2}}\simeq2.9\times10^5~{\rm rad~m^{-2}}\xi\epsilon_{B,-5}^{1/2}M_{0,\odot}^{11/4} E_{51}^{-5/4}t_{10{\rm yr}}^{-7/2},
\ee
and the RM variation is
\be
\frac{d{\rm RM}}{dt}\simeq-\frac{e^3}{2\pi m_e^2c^4}\frac{3^{1/2}7\xi\epsilon_B^{1/2}M^{3/2}}{8\pi\mu_e m_pv^{5/2}t^{9/2}}\simeq1.0\times10^5~{\rm rad~m^{-2}yr^{-1}}\xi\epsilon_{B,-5}^{1/2}M_{0,\odot}^{11/4} E_{51}^{-5/4}t_{10{\rm yr}}^{-9/2}.
\ee
Here a relatively small value of $\epsilon_B\sim10^{-5}$ is adopted considering that the unshocked ejecta is dominated. In this case, the amplification effect of the magnetic field is relatively weak due to the lack of interaction between the gas and the magnetic field. 

II) Shocked ionized ejecta: 
When the SN ejecta propagates in the ISM, a reverse shock and a forward shock will be generated in the SN ejecta and the ISM, respectively. 
The shocks transfer the kinetic energy to the thermal energy, leading to the ejecta medium heated.
If the SN ejecta is neutral and recombined in the free-expansion phase, only electrons in the shocked ejecta could contribute to the DM and RM \cite{piro18}. 
Notice that the contributions of both the DM and RM from the shocked ISM is orders of magnitude smaller than those of the ejecta in the free-expansion phase, thus, we only focus on the former in the following discussion.  We consider that the forward shock has a velocity of $v_b$ at a radius of $R_b$ and the reverse shock has a velocity of $v_r$ at a radius of $R_r$. The contact discontinuity between the SN ejecta and ISM is at the radius of $R_c$. The ISM has a density of $n$. In the rest frame of the unshocked ejecta just ahead of it, the velocity of the reverse shock is $\tilde v_r$.
Following the previous work \cite{piro18}, we calculate both the DM and RM of the shocked SN ejecta as follows.
The DM contributed by the shocked ejecta is
\be
{\rm DM}=\int_{R_r}^{R_c}n_rdr\simeq n_r(R_c-R_r).
\ee
The thickness of the shocked ejecta is roughly 
\be
R_c-R_r\simeq0.434\fraction{t}{t_{\rm ST}}{5/2}R_{\rm ST}\simeq4.7\times10^{-4}~{\rm pc}~M_{0,\odot}^{-7/4}E_{51}^{5/4}n_0^{1/2}t_{10{\rm yr}}^{5/2},
\ee
where $R_{\rm ST}\equiv2.2~{\rm pc}~M_{0,\odot}^{1/3}n_0^{-1/3}$ is the Sedov--Taylor radius, $t_{\rm ST}\equiv210~{\rm yr}~M_{0,\odot}^{5/6}E_{51}^{-1/2}n_0^{-1/3}$ is the Sedov--Taylor timescale, $n=n_0~{\rm cm^{-3}}$ is the electron number density in the ISM. Considering the pressure continuity across the contact discontinuity, $\rho_r\tilde v_r^2\sim\rho_bv_b^2$, where $\rho_r$ and $\rho_b$ are the mass density of the reverse and forward shocks, respectively, the electron number density in the reverse shock is
\be
n_r=4n\fractionz{\mu}{\mu_e}\fraction{v_b}{\tilde v_r}{2}\simeq3.77n\fractionz{\mu}{\mu_e}\fraction{t}{t_{\rm ST}}{-3},
\ee
where $v_b\simeq1.37v_{\rm ST}$ and $\tilde v_r\simeq1.41(t/t_{\rm ST})^{3/2}v_{\rm ST}$ for $t\ll t_{\rm ST}$ \cite{McKee95}, and $v_{\rm ST}\equiv R_{\rm ST}/t_{\rm ST}$. Therefore, the DM contributed by the shocked ejecta is \cite{piro18}
\be
{\rm DM}\simeq16.6~{\rm pc~cm^{-3}}~\mu\mu_e^{-1}M_{0,\odot}^{3/4}E_{51}^{-1/4}n_0^{1/2}t_{10{\rm yr}}^{-1/2},
\ee
and the DM variation is
\be
\frac{d{\rm DM}}{dt}\simeq-0.83~{\rm pc~cm^{-3}yr^{-1}}~\mu\mu_e^{-1}M_{0,\odot}^{3/4}E_{51}^{-1/4}n_0^{1/2}t_{10{\rm yr}}^{-3/2},
\ee
In the reverse shock, the magnetic field is given by $B^2/8\pi\sim\epsilon_B\rho_r\tilde v_r^2/2$, leading to
\be
B\sim\left(4\pi\epsilon_B\rho_r\right)^{1/2}\tilde v_r\simeq1.37\left(16\pi\epsilon_B\mu m_pn\right)^{1/2}v_{\rm ST},
\ee
where $v_b\simeq1.37v_{\rm ST}$ and $\rho_r\tilde v_r^2\sim\rho_bv_b^2$ are used in the last equation. Therefore, the RM contributed by the shocked ejecta is
\be
{\rm RM}\simeq5.7\times10^4~{\rm rad~m^{-2}}~\mu^{3/2}\mu_e^{-1}\epsilon_{B,-1}^{1/2}M_{0,\odot}^{1/4}E_{51}^{1/4}n_0t_{10{\rm yr}}^{-1/2},
\ee
and the RM variation is
\be
\frac{d{\rm RM}}{dt}\simeq-2.8\times10^3~{\rm rad~m^{-2}}~\mu^{3/2}\mu_e^{-1}\epsilon_{B,-1}^{1/2}M_{0,\odot}^{1/4}E_{51}^{1/4}n_0t_{10{\rm yr}}^{-3/2}.
\ee

The scaling law of the evolution of both DM and RM contributed by the SN ejecta in the free-expansion phase is summarized in Table \ref{table}. 
Finally, we discuss the foreground contribution of the DM. Different from the observed RM of FRB 20121102A which should be mainly contributed by the local plasma near the FRB source, the total host DM is composed of the contributions from both the host ISM and the local plasma. 
If the long-term DM evolution is attributed to the expansion of the SN ejecta as discussed above, the foreground host ISM contribution could be estimated as follows.
We define the foreground host ISM contribution as ${\rm DM}_0$. For the scenario of the expanding SNR in the free-expansion phase, one has $({\rm DM}-{\rm DM}_0)\propto t_{\rm SNR}^{-\beta}$, leading to
\be
t_{\rm SNR}=\beta\left|\frac{{\rm DM}-{\rm DM}_0}{d{\rm DM}/dt}\right|.\label{tDM}
\ee
According to Eq.(\ref{tRM}) and Eq.(\ref{tDM}), the DM foreground contribution could be written
\be
{\rm DM}_0={\rm DM}-\frac{\alpha}{\beta}\left|\frac{dt}{d\ln{\rm RM}}\right|\left|\frac{d{\rm DM}}{dt}\right|.
\ee
We take $|dt/d\ln{\rm RM}|\sim10~{\rm yr}$, $|d{\rm DM}/dt|\sim0.43~{\rm pc~cm^{-3}yr^{-1}}$ (for the long-term evolution from 2012 to 2023), ${\rm DM}\sim 203~{\rm pc~cm^{-3}}$, $\alpha/\beta\sim(4/7-1)$. Thus, the contribution from the host ISM is ${\rm DM}_0\sim200~{\rm pc~cm^{-3}}$.

\begin{table*}
\begin{center}
    \caption{Summary of different scenarios for SN in the free-expansion phase }
    \begin{tabular}{cccc}
    \hline\hline
    Different ionization scenarios & Different origin of $B$ field & RM & DM\\
    \hline
    \multirow{2}{*}{Fully ionized ejecta} & Magnetic freezing & $t^{-2}$ & $t^{-3}$\\
     & Energy-equipartition & $t^{-2}$ & $t^{-7/2}$\\
    Shocked ionized ejecta & Energy-equipartition  & $t^{-1/2}$    &$t^{-1/2}$\\
    \hline\hline
    \end{tabular}
    \label{table}
\end{center}
\end{table*}

\subsection{Extra DM from Pair Injection by Enhanced Wind}\label{subsec_pair}

As pointed out in Section \ref{sec_snr}, both the DM and RM will show a power-law decay in the free-expansion phase of an SNR. However, the DM evolution of FRB 20121102A is relatively shallow from 2015 to 2020, which significantly deviates from a power-law decay. We propose that such an extra DM might be contributed by pair injection from the central engine of the FRB source. Meanwhile, since the pair plasma cannot contribute to the RM due to the symmetry of electrons and positrons, the RM evolution is still dominated by the SN ejecta, which is consistent with the observed RM evolution of FRB 20121102A. The pair injection might be from the enhanced wind caused by the dissipation of the magnetic energy of the magnetosphere in the activity phase. 

\begin{figure}
    \centering
    \includegraphics[width = 0.55\linewidth, trim = 100 50 100 50, clip]{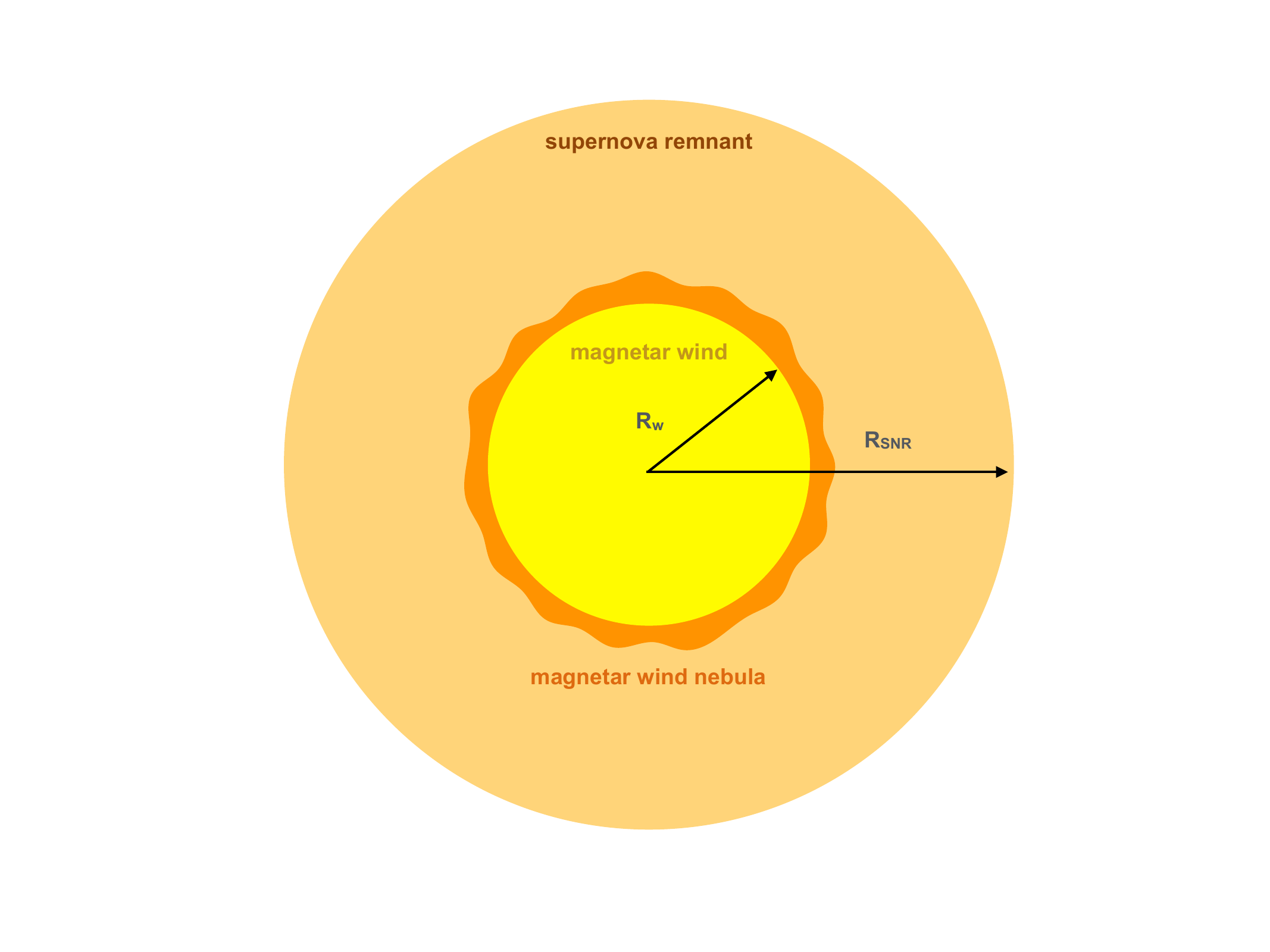}
    \caption{Schematic configuration showing the main regions of focus for the SNR and the magnetar wind nebula. The yellow region corresponds to the unshocked magnetar wind, the orange region corresponds to the shocked magnetar wind (``magnetar wind nebula'') and the cantaloupe-color region corresponds to the SNR.
    Notice that the forward shock region in the SNR is not shown in this figure.}\label{dmrm}
\end{figure}

Before the enhanced wind generation, we first consider that the persistent pair wind caused by the magnetar spindown has a kinetic luminosity of $L_w$ and interacts with the SN ejecta with a mass of $M$, a kinetic energy of $E$, a velocity of $v$ and a density of $\rho$. A mass of $M_s$ is swept up into a thin shell characterized by a radius of $R_w$. The equations for the shell motion, the energy conservation in the wind bubble, and the mass conservation for the shell can be written by \cite{Ostriker71, Chevalier77}
\be
M_s\frac{d^2R_w}{dt^2}=4\pi R_w^2\left[P_c-\rho\left(\frac{dR_w}{dt}-v\right)^2\right],~~~\frac{d(4\pi R_w^4P_c)}{dt}=L_wR_w,~~~\frac{dM_s}{dt}=4\pi R_w^2\rho\left(\frac{dR_w}{dt}-\frac{R_w}{t}\right).
\ee
where the density is $\rho=3M/[4\pi(vt)^3]$. The mass of the swept-up shell is solved as $M_s=M[R_w/(vt)]^3$.
The radius $R_w$ of the ``magnetar wind nebula'' at the time $t$ could be solved \cite{Chevalier77}
\be
R_w\simeq\left(\frac{125}{99}\frac{v^3L_w}{M}\right)^{1/5}t^{6/5}.\label{Rw1}
\ee
Using $v=(2E/M)^{1/2}$ and $R_{\rm SNR}=vt$, the ratio of the radius between the magnetar wind nebula and the SNR is
\be
\frac{R_w}{R_{\rm SNR}}\simeq0.9\fraction{L_wt}{E}{1/5}\sim \fraction{E_w}{E}{1/5},\label{Rw2}
\ee
where $E_w\sim L_wt$ denotes the kinetic energy of the persistent pair wind during time $t$. 

When the magnetar magnetosphere becomes active, an enhanced wind will be generated and reach $R_w$, leading to an extra DM contribution.
In order to generate a DM variation of $\Delta{\rm DM}$, the total number of the pair particles is required to be
\be
\Delta N\sim4\pi R_w^2\gamma\Delta {\rm DM},
\ee
where $\gamma$ is the Lorentz factor of the pair electrons in the enhanced wind nebula. Due to the relativistic motion of the pair electrons, the DM would be suppressed by a factor of $\gamma$.
The total energy from the pair injection is 
\be
\Delta E\simeq \Delta N\gamma m_ec^2=4\pi m_ec^2\gamma^2 R_w^2\Delta{\rm DM}\simeq3.0\times10^{48}~{\rm erg}\gamma_1^2R_{w,-2,{\rm pc}}^2\Delta {\rm DM},
\ee
where $R_{w,-2,{\rm pc}}=R_w/(0.01~{\rm pc})$ and $\Delta {\rm DM}$ is in the unit of ${\rm pc~cm^{-3}}$.
We notice that $\Delta E$ is of the order of the magnetic energy of the magnetosphere. Considering that the extra DM lasts a few years, $\Delta t\sim(1-10)~{\rm yr}$, the luminosity of the enhanced wind is estimated as $L_{w,\rm en}\sim\Delta E/\Delta t\simeq9.5\times10^{39}~{\rm erg~s^{-1}}\gamma_1^2R_{w,-2,{\rm pc}}^2\Delta {\rm DM}\Delta t_{10{\rm yr}}^{-1}$. 
On the other hand, according to Eq.(\ref{Rw1}) or Eq.(\ref{Rw2}), the luminosity of the persistent wind is $L_w\simeq4.5\times10^{37}~{\rm erg~s^{-1}}M_{0,\odot}^{5/2}E_{51}^{-3/2}R_{w,-2,{\rm pc}}^5t_{10{\rm yr}}^{-6}$. Therefore, the luminosity of the enhanced wind is amplified by a factor of 
\be
\frac{L_{w,{\rm en}}}{L_w}\sim210\gamma_1^2R_{w,-2,{\rm pc}}^{-3}M_{0,\odot}^{-5/2}E_{51}^{3/2}\Delta {\rm DM}\Delta t_{10{\rm yr}}^{-1}t_{10{\rm yr}}^{6},
\ee
compared to that of the previous persistent wind.

Finally, the DM significantly falls back to the power-law decay after 2022. 
The reason might be due to the faster expansion of the magnetar wind nebula.
According to Eq.(\ref{Rw2}), after the extra pair injection by the enhanced magnetar wind, the radius of the magnetar wind nebula would reach $R_{w,f}$ with
\be
\frac{R_{w,f}}{R_w}\sim\fraction{L_{w,{\rm en}}\Delta t}{L_w t}{1/5}\sim2.9\gamma_1^{2/5}R_{w,-2,{\rm pc}}^{-3/5}M_{0,\odot}^{-1/2}E_{51}^{3/10}\Delta {\rm DM}^{1/5}t_{10{\rm yr}},
\ee
for the above typical parameters. The extra DM in the magnetar wind nebula is $\Delta{\rm DM}\propto \delta n_{e,w}R_w\propto R_w^{-2}$ assuming that the extra electron number density in the magnetar nebula satisfies $\delta n_{e,w}\propto R_w^{-3}$. Thus, when the enhanced wind finishes, the extra DM contribution would drop by a factor of $(R_{w,f}/R_w)^{-2}\sim10$, rendering the SN ejecta to dominate the DM contribution in the following evolution.
At last, the possible 157-day active window may suggest a possible binary system \cite{ioka20}, which may introduce modulation of DM and RM \cite{wang22,yang23}. Such effects are not evident from the data.

\clearpage
\bibliographystyle{naturemag}

\subsection{Data availability}
Observations log of FRB 20121102A from Mar. 2020 to Apr. 2023 and all relevant data for the 555 detected burst events are summarized in the manuscript Supplementary Table. Observational data are available from the FAST archive\footnote{http://fast.bao.ac.cn} one year after data-taking, following the FAST data policy. Due to the large data volume for these observations, interested users are encouraged to contact the corresponding author to arrange the data transfer.

\subsection{Code availability}
Computational programs for the FRB121102 burst analysis and observations reported here are available at https://github.com/NAOC-pulsar/PeiWang-code. Other standard data reduction packages are available at their respective websites:\\
PRESTO: https://github.com/scottransom/presto\\
DSPSR: http://dspsr.sourceforge.net\\
PSRCHIVE: http://psrchive.sourceforge.net

\clearpage
\section*{Supplementary Table}
\renewcommand{\baselinestretch}{1.0}
\selectfont
\noindent
\EXTTAB{tab:obslog}~1: Observations of FRB 20121102A from Mar. 2020 to Apr. 2023. \\%
\EXTTAB{tab:bursttab}~2: The properties of 555 bursts of FRB 20121102A measured with FAST. \\%

\setcounter{figure}{0}
\setcounter{table}{0}
\captionsetup[table]{name={\bf Supplementary Table} }
\setlength{\tabcolsep}{2.5mm}{
\renewcommand\arraystretch{1.1}
\scriptsize
\begin{longtable}{c c c c c c c c c}%
\caption{Observations of FRB 20121102A from Mar. 2020 to Apr. 2023.}
\\
\hline%
\hline%
Observational Date & MJD$_{start}$ & MJD$_{end}$ & Telescope & Frequency & Duration & Number of & Burst Rate$^{\#}$ \\%
(\textit{UT, YYYYMMDD})& & & & (\textit{GHz}) & (\textit{hour}) & Detections  & (\textit{hour$^{-1}$}) \\%
\hline%
\endhead%
\hline%
\endfoot%
\hline%
\endlastfoot%

20200312 & 58920.397175511 & 58920.522175511 & & & 3.0 & 0 & 0.0 \\%
20200313 & 58921.397175511 & 58921.522175511 & & & 3.0 & 0 & 0.0 \\%
20200314 & 58922.397175511 & 58922.522175511 & & & 3.0 & 0 & 0.0 \\%
20200407 & 58946.315972222 & 58946.441838625 & & & 3.0 & 0 & 0.0 \\%
20200516 & 58985.215578704 & 58985.337642271 & & & 2.9 & 0 & 0.0 \\%
20200601 & 59001.188842177 & 59001.313842177 & & & 3.0 & 0 & 0.0 \\%
20200607 & 59007.215948866 & 59007.340948866 & & & 3.0 & 0 & 0.0 \\%
20200729 & 59058.998553033 & 59059.040219700 & & & 1.0 & 0 & 0.0 \\
\end{longtable}}
\begin{tablenotes}
\item[\#] $\#$ The burst rate is assumed to be a constant throughout the day, defined as the number of detections divided by the duration of time. The Burst rate is zero if no burst from a date was detected in the duration time.
\end{tablenotes}
\clearpage

\setcounter{figure}{1}
\setcounter{table}{1}
\captionsetup[table]{name={\bf Supplementary Table} }
\setlength{\tabcolsep}{0.8mm}{
\renewcommand\arraystretch{2.0}
\scriptsize
\begin{longtable}{c c c c c c c c c c c c}%
\caption{The properties of 555 bursts of FRB 20121102A measured with FAST.}
\\
\hline%
\hline%
Burst & MJD $_{bary}^{a)}$ & DM$^{b)}$ & FWHM$^{c)}$ & W$_{eq}^{d)}$ & Cntrl freq. & Bandwidth$^{e)}$ & Peak\ flux & Fluence & Energy$_{\Delta \nu}^{f)}$ & Energy$_{1.25 GHz}^{g)}$ & Morph.$^{h)}$ \\%
\textit{ID}&(\textit{at infinite freq.}) & (\textit{pc\ cm$^{-3}$}) & (\textit{ms}) & (\textit{ms}) & (\textit{MHz}) & (\textit{MHz}) & (\textit{mJy}) & (\textit{mJy\ ms})  & (\textit{$\times 10^{37}$ erg}) & (\textit{$\times 10^{37}$ erg}) & (\textit{type}) \\%
\hline%
\endhead%
\hline%
\endfoot%
\hline%
\endlastfoot%

1 & 59806.980550710 & 553.9 $^{+0.22}_{-0.22}$ & 1.9 $^{+0.4}_{-0.4}$ & 8.9 $^{+0.4}_{-0.4}$ & 1330 & 320 $^{+32}_{-32}$ $^{(\#1)}$ & 10.7 $^{+0.1}_{-0.1}$ & 94.63 $^{+0.05}_{-0.05}$ & 3.26 $^{+0.004}_{-0.004}$ & 9.56 $^{+0.01}_{-0.01}$ &  NB-MC-FD   \\%
2 & 59806.980553938 & 551.52 $^{+0.24}_{-0.24}$ & 1.1 $^{+0.4}_{-0.4}$ & 19.9 $^{+2.3}_{-2.3}$ & 1110 & 120 $^{+12}_{-12}$ $^{(\#1)}$ & 11.4 $^{+0.1}_{-0.1}$ & 227.06 $^{+0.28}_{-0.28}$ & 2.94 $^{+0.008}_{-0.008}$ & 14.04 $^{+0.06}_{-0.06}$ &  NB-SC   \\%
3 & 59806.985587007 & 552.93 $^{+0.39}_{-0.39}$ & 1.5 $^{+0.4}_{-0.4}$ & 8.5 $^{+0.7}_{-0.7}$ & 1300 & 400 $^{+40}_{-40}$ $^{(\#1)}$ & 7.5 $^{+0.0}_{-0.0}$ & 64.10 $^{+0.05}_{-0.05}$ & 2.76 $^{+0.006}_{-0.006}$ & 7.24 $^{+0.01}_{-0.01}$ &  NB-SC   \\%
4 & 59806.992097825 & 552.86 $^{+0.45}_{-0.45}$ & 0.9 $^{+0.4}_{-0.4}$ & 14.7 $^{+1.8}_{-1.8}$ & 1110 & 120 $^{+12}_{-12}$ $^{(\#1)}$ & 11.0 $^{+0.1}_{-0.1}$ & 161.41 $^{+0.18}_{-0.18}$ & 2.09 $^{+0.005}_{-0.005}$ & 9.98 $^{+0.04}_{-0.04}$ &  NB-SC   \\%
5 & 59812.082410260 & 551.94 $^{+0.35}_{-0.35}$ & 1.5 $^{+0.4}_{-0.4}$ & 4.5 $^{+0.3}_{-0.3}$ & 1130 & 240 $^{+24}_{-24}$ $^{(\#1)}$ & 14.2 $^{+0.1}_{-0.1}$ & 64.69 $^{+0.05}_{-0.05}$ & 1.67 $^{+0.003}_{-0.003}$ & 5.66 $^{+0.01}_{-0.01}$ &  NB-SC   \\%
6 & 59812.082416337 & 553.74 $^{+0.4}_{-0.4}$ & 1.4 $^{+0.6}_{-0.6}$ & 8.9 $^{+0.5}_{-0.5}$ & 1270 & 200 $^{+20}_{-20}$ $^{(\#1)}$ & 14.7 $^{+0.1}_{-0.1}$ & 130.37 $^{+0.09}_{-0.09}$ & 2.81 $^{+0.005}_{-0.005}$ & 10.41 $^{+0.02}_{-0.02}$ &  NB-SC   \\%
7 & 59812.093667546 & 552.47 $^{+0.27}_{-0.27}$ & 1.2 $^{+0.4}_{-0.4}$ & 8.3 $^{+0.8}_{-0.8}$ & 1340 & 320 $^{+32}_{-32}$ $^{(\#1)}$ & 10.5 $^{+0.1}_{-0.1}$ & 87.51 $^{+0.10}_{-0.10}$ & 3.02 $^{+0.009}_{-0.009}$ & 8.84 $^{+0.02}_{-0.02}$ &  NB-SC   \\
\end{longtable}}
\begin{tablenotes}
\item[a)] $a)$ Arrival time of burst peak at the solar system barycenter, after correcting to the infinite frequency. \\%
\item[b)] $b)$ DM obtained from the best burst alignment, calculated using the \textit{DM-Power} algorithm (https://github.com/hsiuhsil/DM-power). A '-' is marked and the DM uses the daily averaged value instead for the bursts where the DM-fit failed.\\%
\item[c)] $c)$ The FWHM is defined as the full width at half maximum of the autocorrelation function (ACF) for a pulse profile, i.e. the measured center value is the width where the pulse fluence is 50$\%$, and the error is obtained by Monte Carlo (MC) fitting: take the averaged value of the pulse profiles,1000 random samples according to the variance of the distribution for the off-pulse background noise.\\%
\item[d)] $d)$ The equivalent width W$_{eq}$ was defined as the width of a rectangular burst that has the same area as the profiles, with the height of peak flux density. \\%
\item[e)] $e)$ Three types of burst bandwidth fit were considered: $^{\#}$1) Boxcar bandwidth of the burst for S/N $\textless$10, a conservative 20$\%$ fractional error is assumed, $^{\#}$2) Gaussian fitting for S/N $\geq$10, uncertainty of measurement was 1$\sigma$, $^{\#}$3) A frequency extended Gaussian fit was used to correct the bandwidth for the bursts were cut off by the upper/lower limit of the observed frequency, 1$\sigma$  measuring uncertainty. \\%
\item[f)] $f)$ The isotropic energy of the bursts was calculated using a partial bandwidth, see text. \\%
\item[g)] $g)$ The isotropic energy of the bursts was replaced by a calculation of the center frequency 1.25 GHz (full bandwidth), see text. \\%
\item[h)] $h)$ Morphological types of the bursts: Broad-Band (BB), Narrow-Band (NB), Single-Components (SC), Multiple-Components (MC), and Frequency-Drifting (FD). \\%
\end{tablenotes}

\end{methods}
\end{document}